\documentclass{aa}
\usepackage{natbib}
\usepackage{txfonts}
\usepackage{hyperref}
\bibpunct{(}{)}{;}{a}{}{,}

\begin{document}

\title{Radiative hydrodynamic modelling and observations of the X-class solar flare on 2011 March 9}
\author{{Michael B. Kennedy}\inst{1}
\and{Ryan O. Milligan}\inst{1,2,3}
\and{Joel C. Allred}\inst{2}
\and{Mihalis Mathioudakis}\inst{1}
\and{Francis P. Keenan}\inst{1}}

\institute{Astrophysics Research Centre, School of Mathematics \& Physics, Queen's University Belfast,  Belfast, UK, BT7 1NN
\and Solar Physics Laboratory, Heliophysics Science Division, NASA Goddard Space Flight Center, Greenbelt, MD 20771, USA
\and Department of Physics, Catholic University of America, 620 Michigan Ave., N.E., Washington, DC 20064}
\date{received/accepted}
\abstract
{}
{We investigated the response of the solar atmosphere to non-thermal electron beam heating using the radiative transfer and hydrodynamics modelling code RADYN. The temporal evolution of the parameters that describe the non-thermal electron energy distribution were derived from hard X-ray observations of a particular flare, and we compared the modelled and observed parameters.}
{The evolution of the non-thermal electron beam parameters during the X1.5 solar flare on 2011 March 9 were obtained from analysis of \textit{RHESSI} X-ray spectra. The RADYN flare model was allowed to evolve for 110 seconds, after which the electron beam heating was ended, and was then allowed to continue evolving for a further 300s. The modelled flare parameters were compared to the observed parameters determined from extreme-ultraviolet spectroscopy.}
{The model produced a hotter and denser flare loop than that observed and also cooled more rapidly, suggesting that additional energy input in the decay phase of the flare is required. In the explosive evaporation phase a region of high-density cool material propagated upward through the corona. This material underwent a rapid increase in temperature as it was unable to radiate away all of the energy deposited across it by the non-thermal electron beam and via thermal conduction. A narrow and high-density ($n_{e} \le 10^{15}$ cm$^{-3}$) region at the base of the flare transition region was the source of optical line emission in the model atmosphere. The collision-stopping depth of electrons was calculated throughout the evolution of the flare, and it was found that the compression of the lower atmosphere may permit electrons to penetrate farther into a flaring atmosphere compared to a quiet Sun atmosphere.}
{}
\titlerunning{RHD Modelling and Observations of a X-class Flare}
\authorrunning{M.B. Kennedy et al.}
\maketitle

\section{Introduction}\label{s:intro}

Non-thermal electrons are thought to be accelerated in the corona during solar flares and lose the majority of their energy in the lower atmosphere through electron-electron Coulomb collisions. A small amount of their energy is lost through electron-ion bremsstrahlung radiation, producing hard X-ray (HXR) photons. If the accelerated electrons are thermalised in the unresolved HXR source region within the observational interval, then the emission is referred to as thick-target. Collisional thick-target models \citep{brow71,huds72,brow73} were originally proposed to explain HXR bursts and have since become the standard model to explain energy transfer into and heating of the low solar atmosphere during the flare impulsive phase.

The hydrodynamics and radiation of a loop subject to thick-target heating was studied by\citet{fish85a,fish85b,fish85c}, who found two distinct types of hydrodynamic response to beam heating based on the non-thermal electron energy flux, referred to as gentle and explosive evaporation. Gentle evaporation occurs when the heating rate in the chromosphere is insufficient to increase the temperature past the peak in the radiative loss function, while explosive evaporation occurs when the heating timescale is shorter than the hydrodynamic expansion timescale. \citet{hawl94} included the non-local thermodynamic equilibrium (NLTE) radiative transfer code MULTI \citep{carl86} in their solar flare models to calculate optically thick transitions from hydrogen, singly ionised calcium, and magnesium, but limitations were imposed on the hydrodynamics of the atmosphere. The RADYN code \citep{carl92,carl94,carl95} was adopted by \citet{abbe99} to study the detailed radiative hydrodynamics of a loop subjected to electron beam heating. Improvements to the code were made by \citet{allr05} to include heating from an injected electron energy spectrum characterised by a double-power-law distribution and time-dependent beam heating parameters. Additional transitions were also included in the treatment of soft X-ray (SXR), extreme-ultraviolet (EUV), and ultra-violet (UV) heating, referred to as radiative back-warming \citep{mach89}. 

There are recent examples in the literature of applications of the code to flare studies. \citet{chen10} employed RADYN to study continuum emissions in white-light solar flares as a function of heliographic angle and energy deposition rate into a sunspot atmosphere. It has also been used to compare the observed and predicted intensity in emission lines and passbands during solar flares. \citet{rubi12} examined the intensity of the H$\alpha$ and Ly$\alpha$ emission lines during several flares and compared the results with the synthetic line intensities from a number of RADYN models. They found that the predicted H$\alpha$ intensity matched with observations to within an order of magnitude and agreed better with the observations than the predicted Ly$\alpha$ intensity. \citet{chen12} employed RADYN to create synthetic UV light curves by convolving the modelled UV continuum with the TRACE 1600\AA\ passband. The modelled emission decayed faster than the observations, though the heating lasted for only 2.5s and the contribution of  \ion{C}{IV} emission lines to the filter were not considered. Radiative hydrodynamic simulations have applications to the study of stellar flares \citep{allr06, kowa13} as there is a greater wealth of optical spectra available for investigation compared to solar flares.

Previous studies involving flare modelling have typically used general parameters that characterise the non-thermal electron energy distribution. However, recent multi-wavelength observations of solar flares from X-ray to optical wavelengths have motivated detailed radiative hydrodynamical modelling of flares with realistic, time-dependent electron beam heating functions derived from HXR observations, such as the detailed study of the X2.2 event on 2011 February 15 by \citet{mill14}. The aims of such studies are to investigate the chromospheric response to electron beam heating, determine which line and continuum transitions dominate the radiative energy losses, and directly compare the model results to optical, UV, and EUV observations of the same event.

In this study we model the atmospheric response to non-thermal electron beam heating using RADYN with the parameters of the injected electron energy spectrum derived from RHESSI HXR observations of a specific flare. The results of the model are then directly compared to physical parameters derived by other instruments, including hydrodynamic variables and radiative output of EUV emission lines and continua observed by the Extreme-ultraviolet Variability Experiment \citep[EVE,][]{wood12}. In Sect. \ref{s:observ} we describe the analysis of the HXR emission used to determine the input parameters to the modelling code, while in Sect. \ref{s:model} the calculation and evolution of the model atmosphere is outlined. Section \ref{s:comp} shows parameters of the flare derived from observations and compares these with the model calculations. A discussion of the results and our conclusions are presented in Sect. \ref{s:conclu}.

\section{Observations}\label{s:observ}

\begin{figure}
\centering
\includegraphics[width=\linewidth]{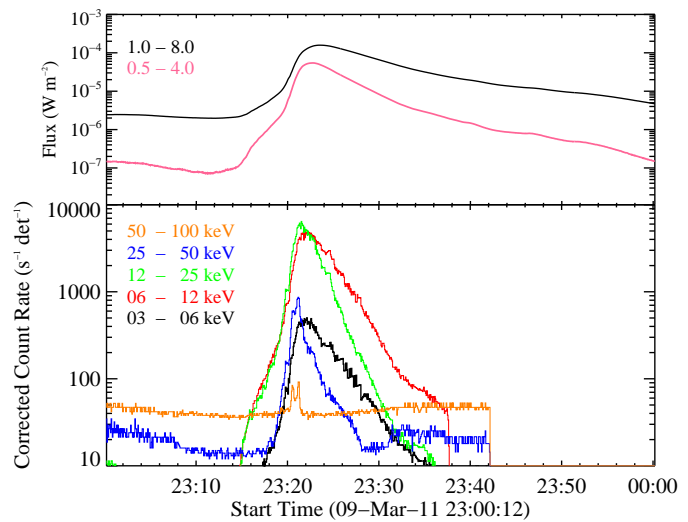} 
\caption{Top panel: GOES light curves from the 1.0 - 8.0 \AA\ (black line) and 0.5 - 4.0 \AA\ (pink line) channels. Bottom panel: RHESSI X-ray light curves in five energy bands.}
\label{f1}
\end{figure}

The event analysed as part of this study was a \textit{Geostationary Orbiting Environmental Satellites} (GOES) X1.5 class flare that occurred in the $\beta \gamma \delta$ NOAA active region 11166 (N08W11) on 2011 March 9 (SOL2011-03-09T23:23). It was observed by instruments onboard the \textit{Solar Dynamics Observatory} \citep[SDO,][]{pesn12} and by the \textit{Reuven Ramaty High Energy Solar Spectroscopic Imager} \citep[RHESSI,][]{lin02} until it entered the night-time portion of its orbit at approximately 23:38 UT. Increased pre-flare SXR emission from thermal plasma was observed starting at 23:15 UT, and increased HXR emission up to energies of 100 keV were detected from 23:20 - 23:22 UT (see Fig. \ref{f1}). Observations from RHESSI were employed to determine the non-thermal electron energy distribution and HXR source area of the flare, with the aim of adopting these parameters as input to the modelling code RADYN. 

\subsection{X-ray imaging}\label{ss:image}

To derive the energy deposition rate into the lower atmosphere, it was necessary to estimate the footpoint area of the event. The area of the footpoint HXR emission was estimated using the imaging capabilities of RHESSI. Initial images of the thermal and non-thermal emissions were made using the Clean algorithm \citep{hurf02} over a number of energy bands. However, for estimates of the HXR footpoint source area, the Pixon algorithm \citep{metc96} was used. In a study of HXR source sizes, \citet{denn09} found that the Pixon algorithm, the maximum-entropy method, and Clean components gave similar estimates of the HXR area. A similar study by \citet{warm13} also found consistent source area estimates among different imaging algorithms.  

\begin{figure*}
\centering
\begin{tabular}{cccc}
\hspace{-0.2in}\includegraphics[width=0.25\linewidth]{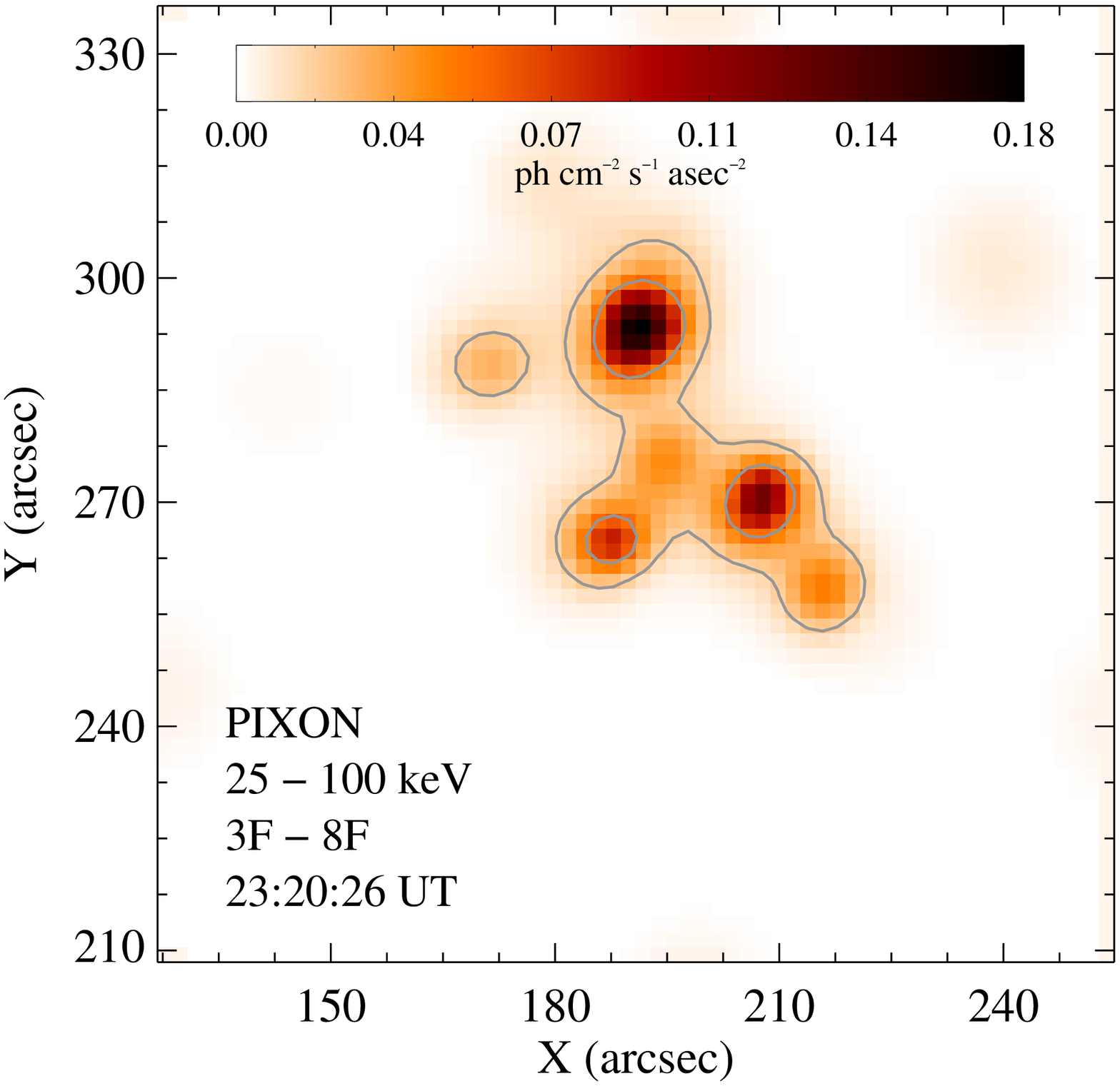} &
\hspace{-0.2in}\includegraphics[width=0.25\linewidth]{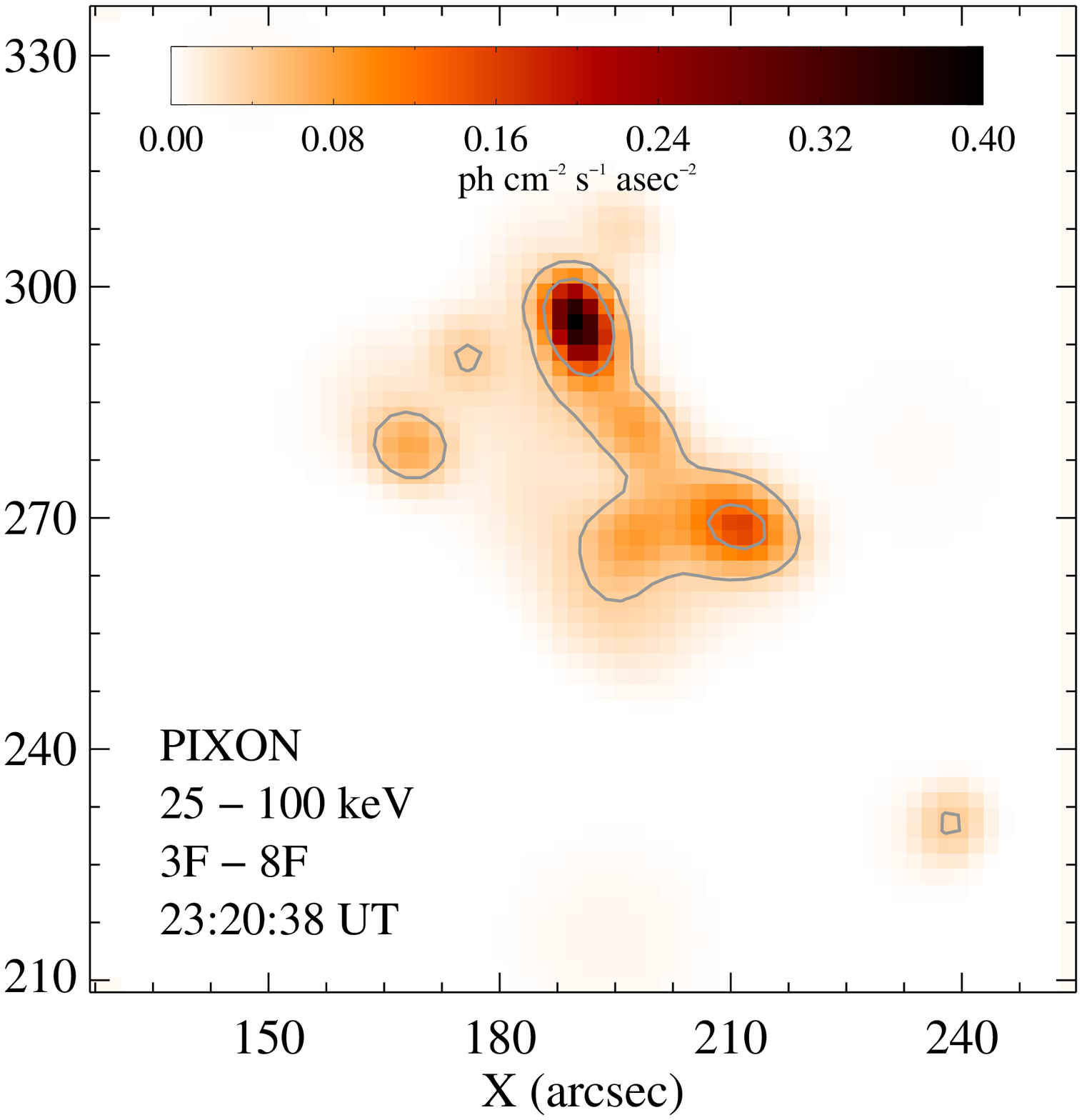} &
\hspace{-0.2in}\includegraphics[width=0.25\linewidth]{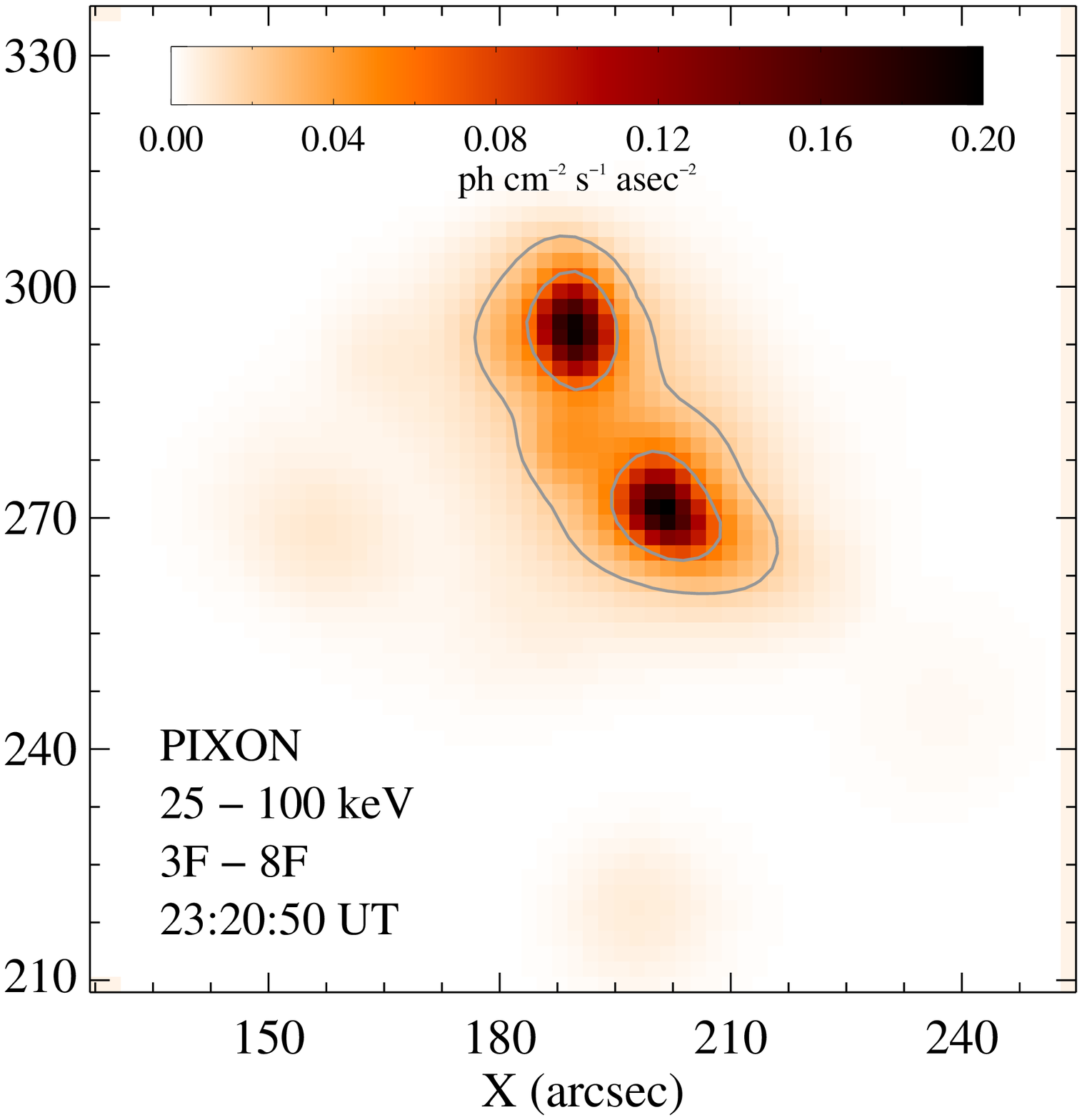} &
\hspace{-0.2in}\includegraphics[width=0.25\linewidth]{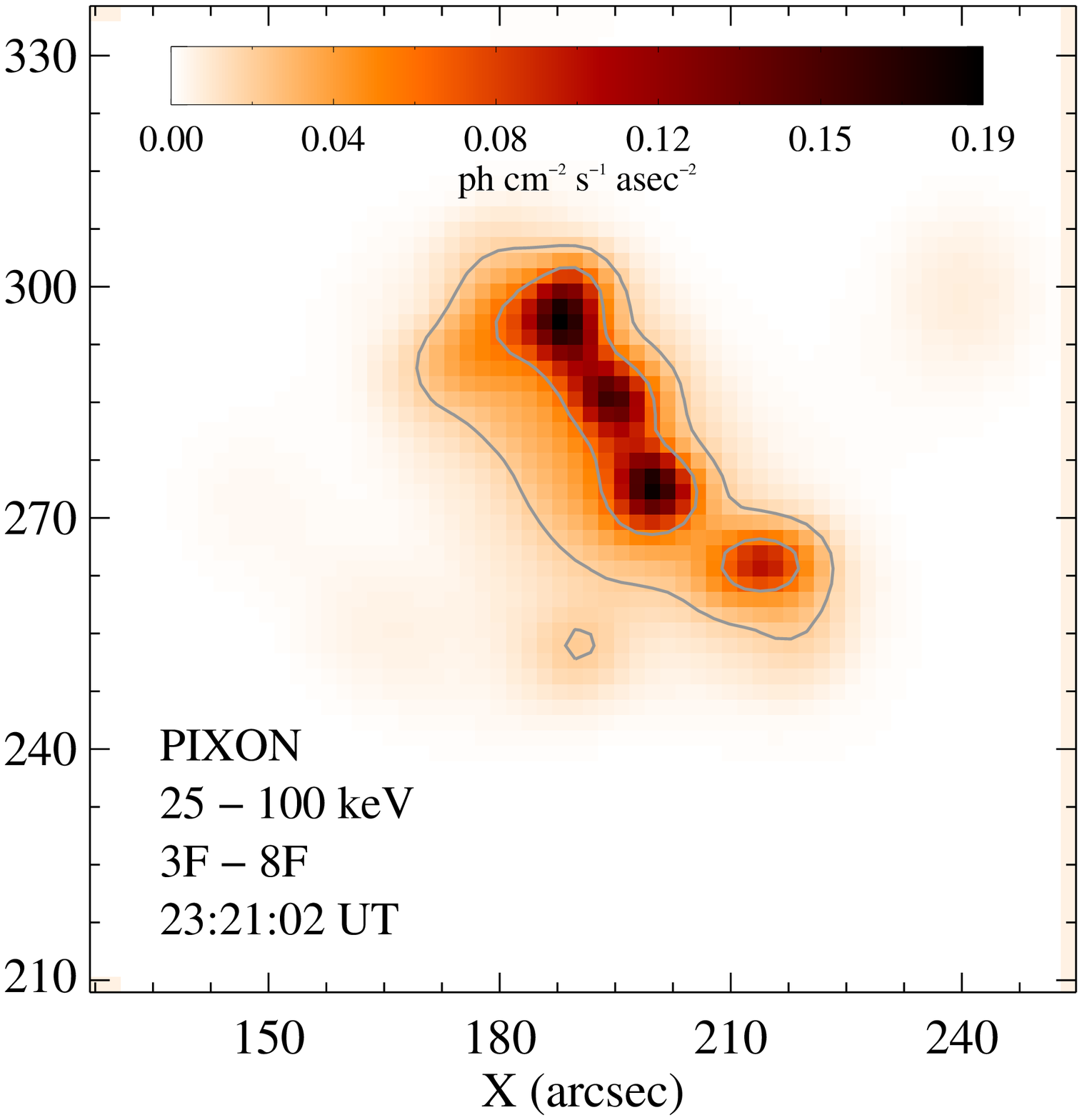} \\
\hspace{-0.2in}\includegraphics[width=0.25\linewidth]{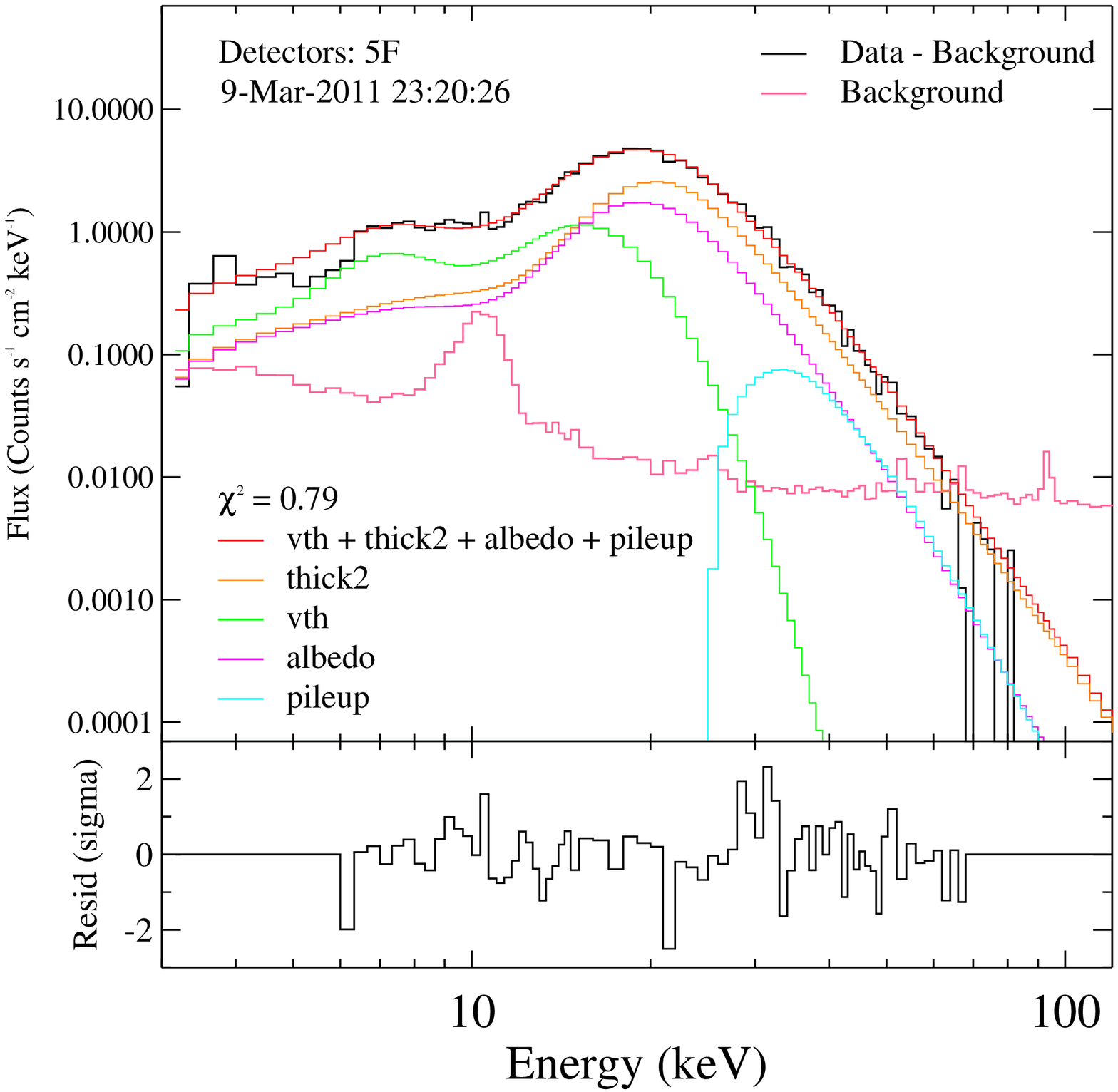} &
\hspace{-0.2in}\includegraphics[width=0.25\linewidth]{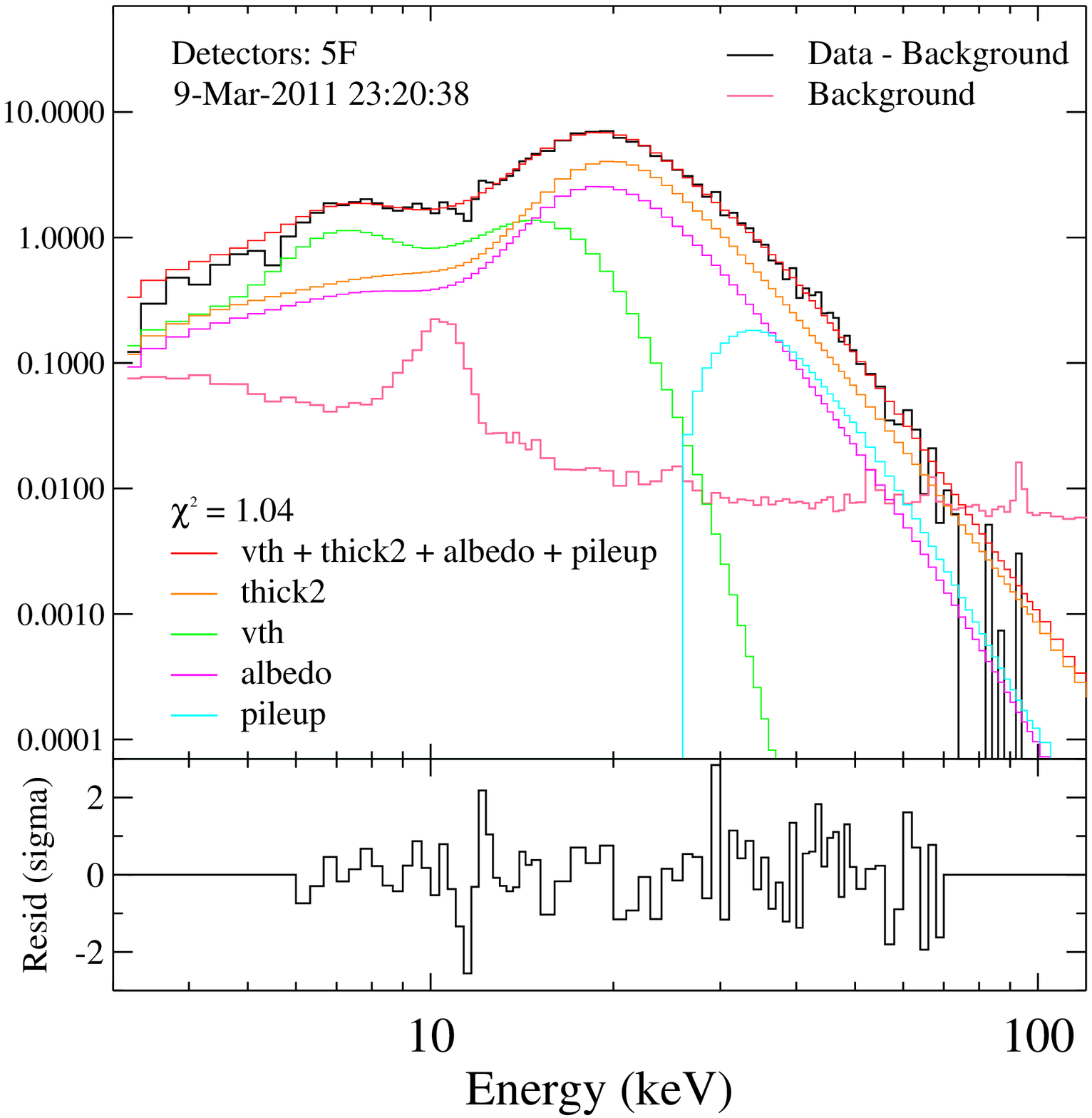} &
\hspace{-0.2in}\includegraphics[width=0.25\linewidth]{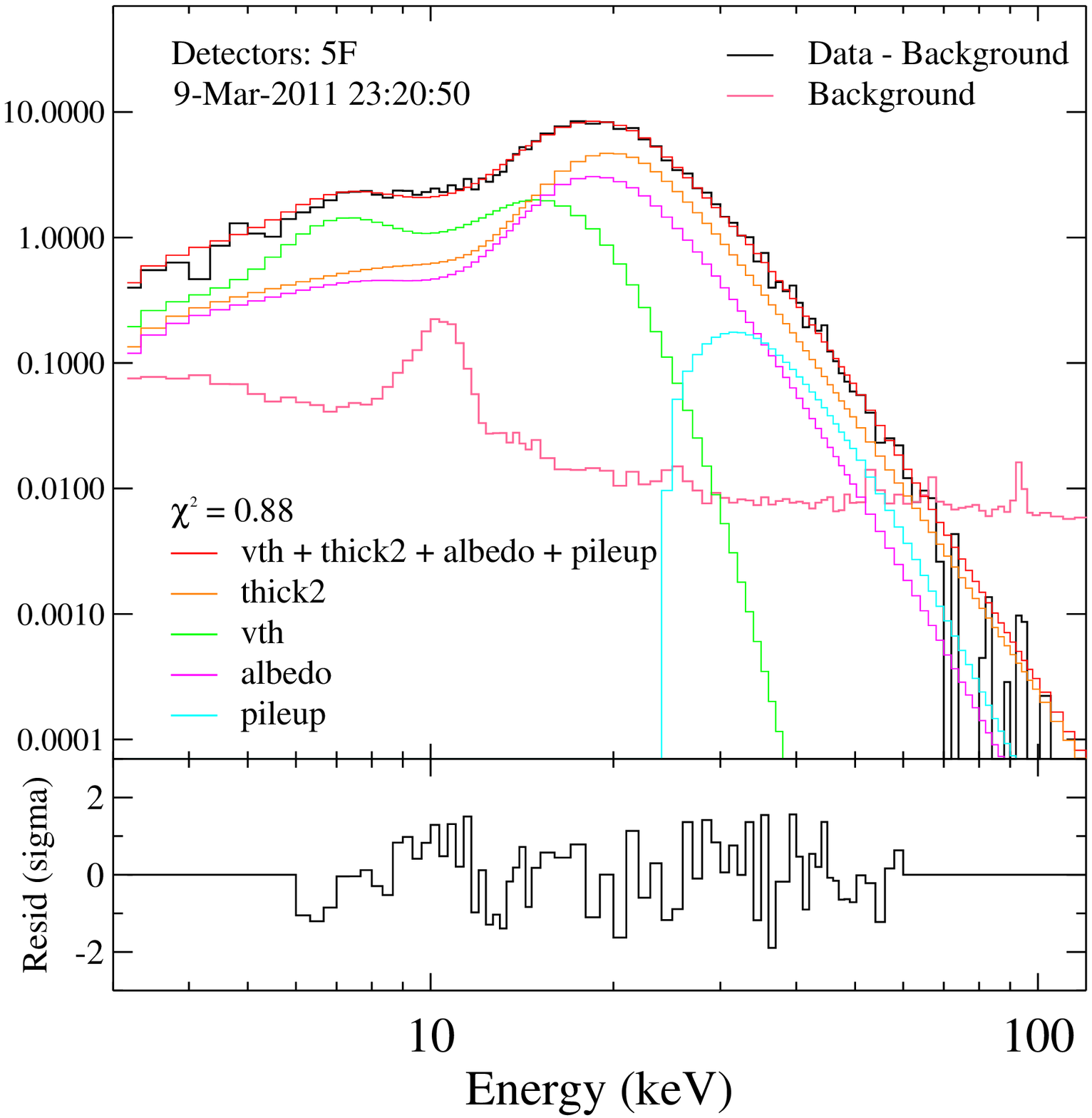} &
\hspace{-0.2in}\includegraphics[width=0.25\linewidth]{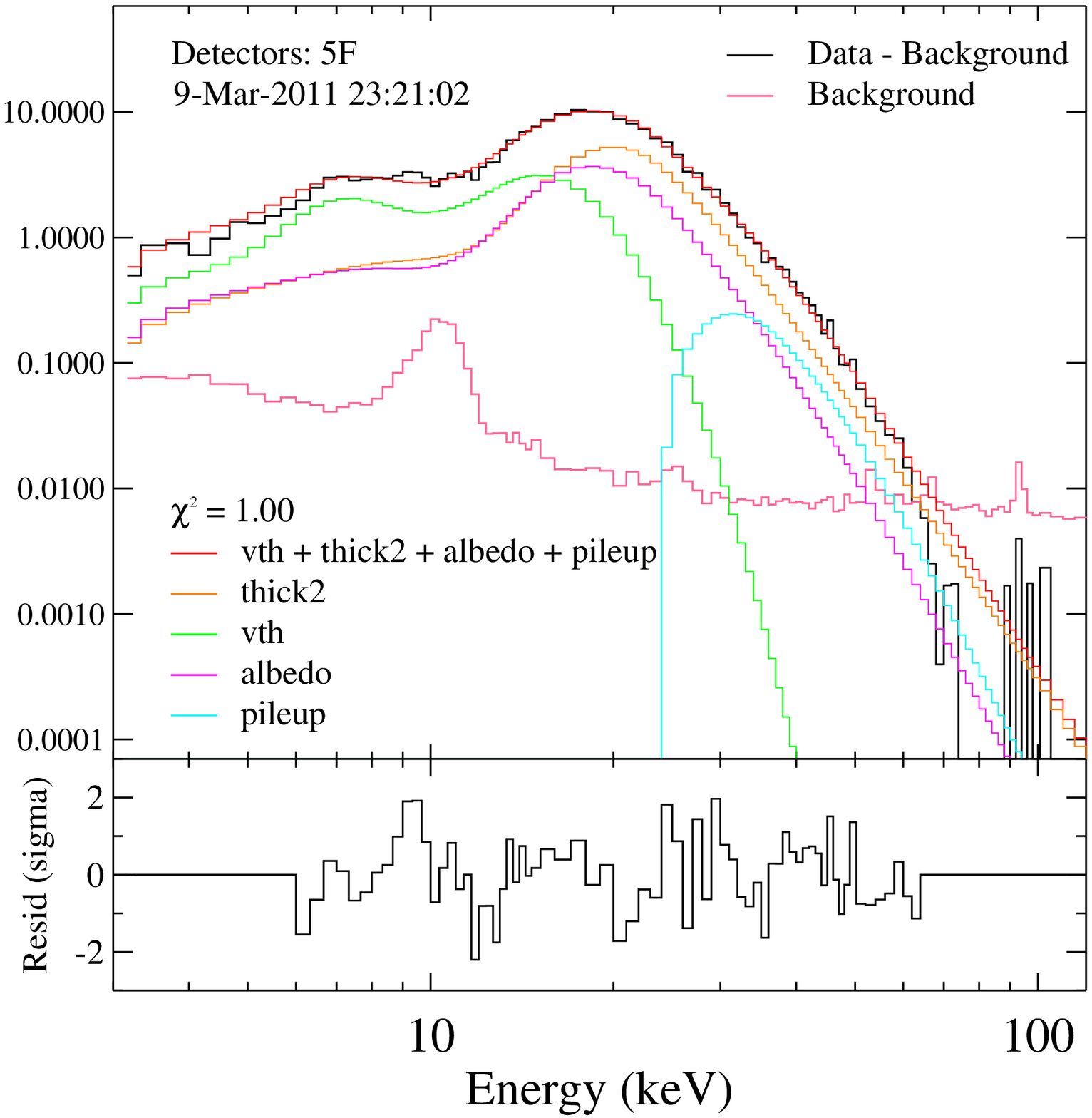} \\
\end{tabular}
\caption{Top row: RHESSI imaging of the flare on 2011 March 09 over the time range UT 23:20:20 - 23:21:08. Images were created over the energy range 25.0 - 100.0 keV using the Pixon algorithm with an integration time of 12 seconds and detectors 3F-8F. The applied colour scaling is based on the highest and lowest data values present in the images displayed. Overplotted in grey are the 10\%\ and 30\% intensity contours. Bottom row: Spectral fits in count flux units made during the same intervals.The background-subtracted data are plotted in black with the background level plotted in pink. The total fit is plotted in red along with the thermal component (green), the thick-target component (orange), albedo component (purple), and pulse pileup correction (cyan). The normalised residuals are shown in each plot in units of sigma.}
\label{f2}
\end{figure*}

Detectors 3F - 8F were employed to create images of the flare over the energy range 25 - 100 keV for a time period of approximately three minutes around the peak in HXR emission, 23:19 to 23:22 UT. The pixel sizes were 2\arcsec\ $\times$ 2\arcsec\ and the imaging intervals were selected to correspond with the spectral fit intervals. Constructed images are shown in Fig. \ref{f2} for four intervals during the impulsive phase. From 23:20:20 UT there are two bright footpoints present until 23:21:00 UT, when several bright sources are apparent in the images. After 23:21:20 UT there is a sharp reduction in counts above 25 keV. The sources present in images after this time are broken up and diffuse. 

The HXR source area in each image was estimated as that contained within a certain intensity threshold. In this case, we adopted 30\% of the highest intensity in each image. The adopted criteria for determining the source area provided estimates that ranged from $8\times10^{17}$ - $2\times10^{18}$ cm$^{2}$ over the time period 23:19:58 to 23:21:20 UT.  For a single interval (23:20:44 - 23:20:56) the area was also estimated by fitting 2D Gaussian profiles to the two footpoint sources visible in the image and using the equation $A = \pi a b $, where $a$ and $b$ are the major and minor axes of the 2D Gaussian. The area derived using this method was found to be $2.4 \times 10^{17}$ cm$^{2}$ for the northern footpoint and $2.2 \times 10^{17}$ cm$^{2}$ for the southern footpoint, approximately half of the value estimated by summing the pixel area within the intensity threshold. This method was not adopted as the sources in each image could not always be adequately represented by a 2D Gaussian profile. The affect of albedo may also have contributed to the estimated footpoint area. \citet{batt11} estimated that albedo could result in increased source areas by up to 30\% for energies between 10 and 100 keV. Given these considerations, the measured footpoint area from HXR imaging should be considered as an overestimate.

\subsection{White-light imaging}\label{ss:wlim}

\begin{figure}
\centering
\includegraphics[width=0.9\linewidth]{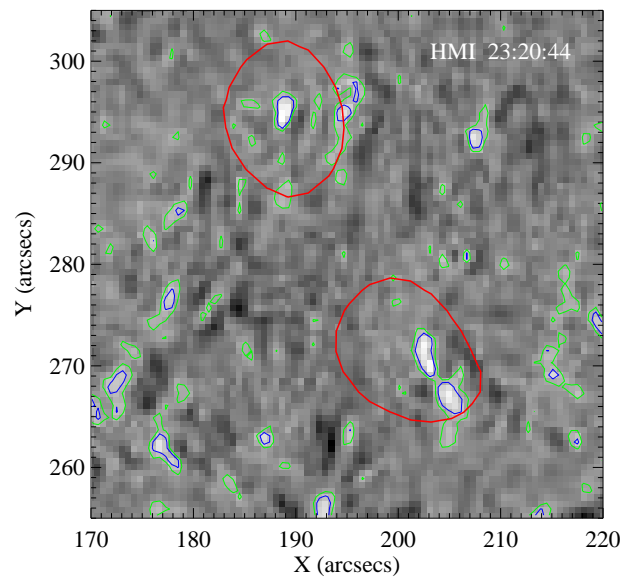}
\caption{HMI difference image at 23:20:44 UT. Overlaid are the 30\% contours (red) from the closest HXR image at 23:20:50 UT and the 30\% and 50\% white-light intensity contours (green and blue).}
\label{f3}
\end{figure}

In addition to the source area estimates from HXR images, continuum intensity observations made by the Helioseismic and Magnetic Imager \citep[HMI,][]{sche12} were examined in an attempt to identify and estimate the size of the optical flare kernels. A mean image was calculated from 23:15:29  - 23:18:29 UT and subtracted from the images taken during the impulsive phase of the flare. Two optical kernels were identified within the area of the HXR footpoints. Figure \ref{f3} displays the HMI difference image at 23:20:44 UT, with the 30\% HXR contours overlaid in red and the 30\% and 50\% WL intensity contours overlaid in green and blue, respectively. A 10\arcsec\ $\times$ 10\arcsec\ region was  centred around each footpoint and the area of the WL emission estimated as that contained within the 50\% intensity contours in these regions. This provided estimates of the WL emitting area to be 7.8 $\times$ 10$^{16}$ to 1.0 $\times$ 10$^{17}$ cm$^{2}$ between 23:20 and 23:22 UT, in agreement with previous studies of flare kernels in optical observations that obtained area estimates in the range $10^{16}$ to $10^{17}$ cm$^{2}$ \citep{kruc11,xu12} and an order of magnitude less than that obtained from HXR imaging.

\subsection{Spectroscopy}\label{ss:spec}

The HXR spectroscopic analysis was performed using disk-integrated RHESSI spectra in \textit{Object Spectral Analysis Executive} (OSPEX). Spectral fits were made individually to six front detectors (1, 4, 5, 6, 8, and 9). The process of fitting to each detector individually was used to verify that the parameters returned from each detector were consistent with the others. The use of a single detector or combined data from a subset of detectors could have introduced a systematic error into the derived parameters of the thermal and non-thermal emissions. This process led to detector 3 being excluded from the analysis as the results obtained were not consistent with the other detectors analysed. An average value of the parameters derived from the six detectors at each interval was adopted, with the error estimated as the standard deviation \citep{mill14}. Spectra were fit in 12-second intervals from approximately 23:20 to 23:21:30 UT with a longer interval on either side to account for the lower count rate. The intervals were chosen to ensure the spectra were not fit across a change of attenuator state and to match with the imaging intervals. Spectral fits for four intervals are shown in Fig. \ref{f2}. Standard corrections for decimation and pulse pileup were applied, and the affect of photospheric albedo \citep{kont06} was corrected for assuming isotropic emission. An isothermal and thick-target bremsstrahlung component were forward-fit to the spectra. The non-thermal emissions were well represented by a single power law over the duration of the flare. Therefore, the electron distribution was characterised by three free parameters: the spectral index ($\delta$), the low-energy cutoff ($E_{c}$) and the electron flux ($F_{0}$), and the thermal component by two free parameters, the temperature ($T$) and emission measure ($EM$). Throughout the duration of the flare, the spectra would be characterised as soft, with values of the spectral index in the range 7 - 8, and the highest emission detected above the background was typically 70 - 80 keV around the flare peak.  As a result of the very soft spectra, there was no apparent break between the thermal and non-thermal emission.

\subsection{Parameter evolution}\label{ss:paramevol}

\begin{figure}
\centering
\includegraphics[width=0.95\linewidth]{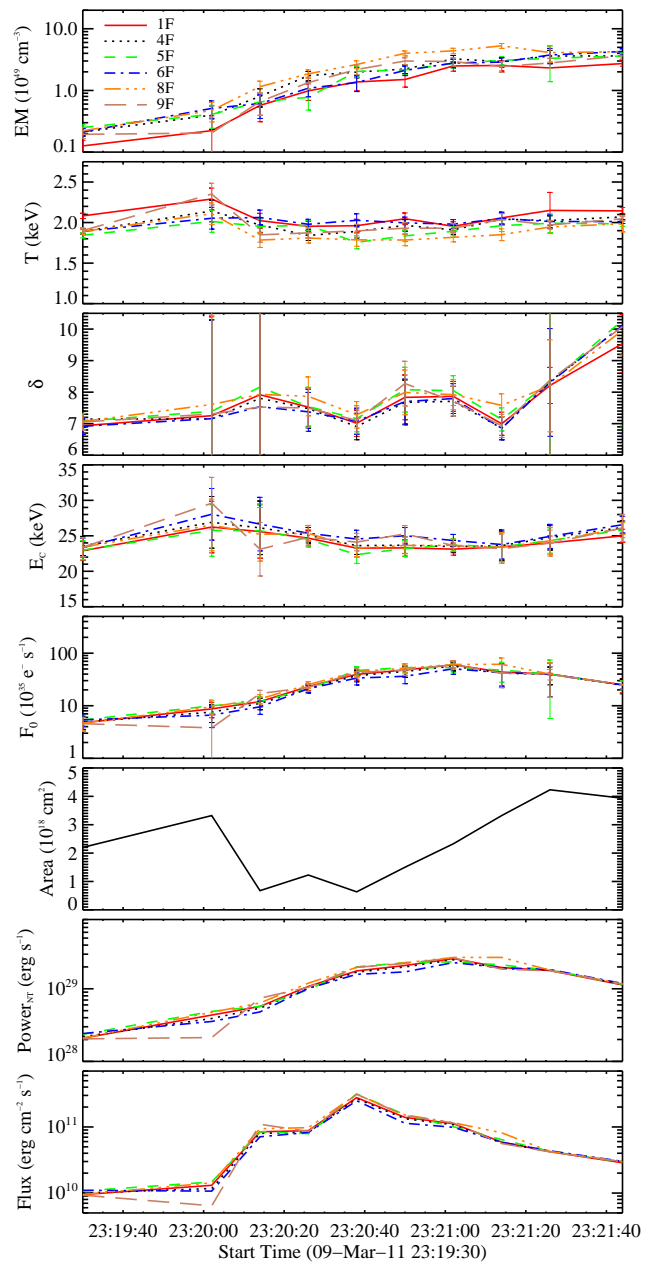} 
\caption{Evolution of the electron beam parameters determined from HXR spectral analysis. The quantities plotted from top to bottom are the emission measure, temperature, spectral index, low energy cut-off, total electron flux, estimated source area, power in non-thermal electrons above the cutoff, and the derived energy deposition rate into the atmosphere. The parameters from detector 1 are plotted as a solid line (red), detector 4 as a dotted line (black), detector 5 as a dashed line (green), detector 6 as a dash-dot-dash line (blue), detector 8 as a dash-dotted line (yellow), and detector 9 as a long dashed line (brown).}
\label{f4}
\end{figure}

The evolution of the parameters derived from HXR spectroscopy and imaging are shown in Fig. \ref{f4}. There was little variation in the value of the low-energy cutoff, and it remained between 22 - 28 keV for the time period studied. The spectral index was soft over the entire flare duration, ranging between 7 and 8, but did become harder at approximately 23:20:30 UT and 23:21:15 UT corresponding to the two peaks in the 50 - 100 keV light curve. Estimates of the HXR source area reached a minimum at the flare peak and then increased with time as the sources became more extended. After 23:21:20 UT, the spectral index steepened and the low-energy cutoff decreased, resulting in the calculated energy content in non-thermal electrons increasing while at the same time the observed counts above 25 keV decreased. The parameters returned from the spectral fits made to individual detectors became discrepant after this time. 

Dividing the power contained in non-thermal electrons above the low-energy cutoff by the HXR source area provided an estimate of the energy deposition rate into the atmosphere. The flux increased from $10^{10}$ erg cm$^{-2}$ s$^{-1}$ at 23:19:30 UT ($t = 0$), reaching $\approx 10^{11}$ erg cm$^{-2}$ s$^{-1}$ after approximately a further 50 seconds, while the highest energy deposition rate of $2.5 \times 10^{11}$ erg cm$^{-2}$ s$^{-1}$ occurred at 23:20:40 UT. While the WL source areas indicated that the energy deposition rate into the atmosphere could be greater than $10^{12}$ erg cm$^{-2}$ s$^{-1}$, these sources would be formed at photospheric heights. If the magnetic field lines converge towards the photosphere, then it may not be appropriate to use them to derive the energy deposition rate of the lower energy electrons stopped higher in the atmosphere. The derived electron parameters were adopted as input to the simulation until 23:21:20 UT. After this time, there did not appear to be compact footpoint sources present in the images. 

\begin{figure*}
\centering
\includegraphics[width=0.93\linewidth]{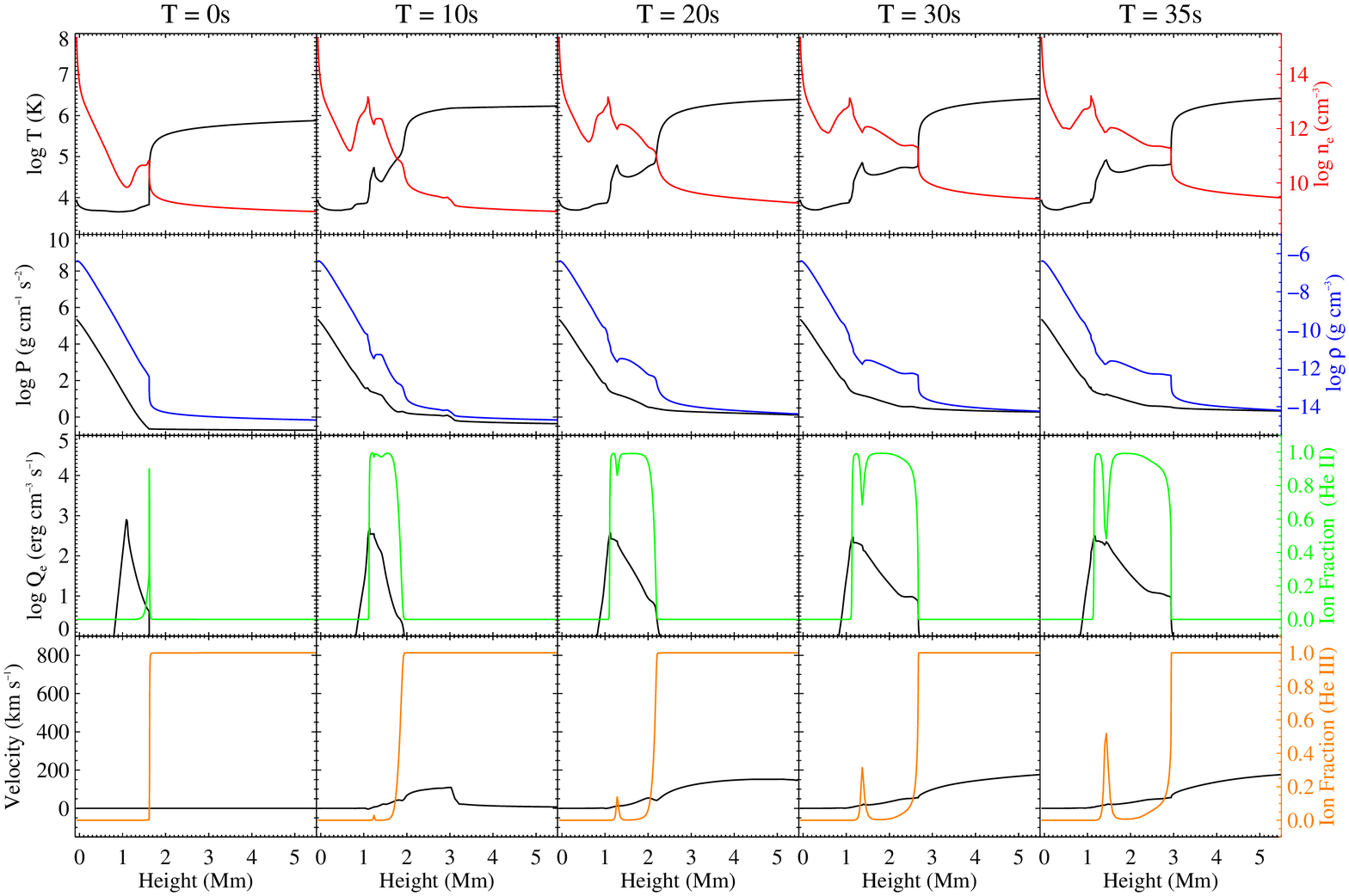}
\includegraphics[width=0.93\linewidth]{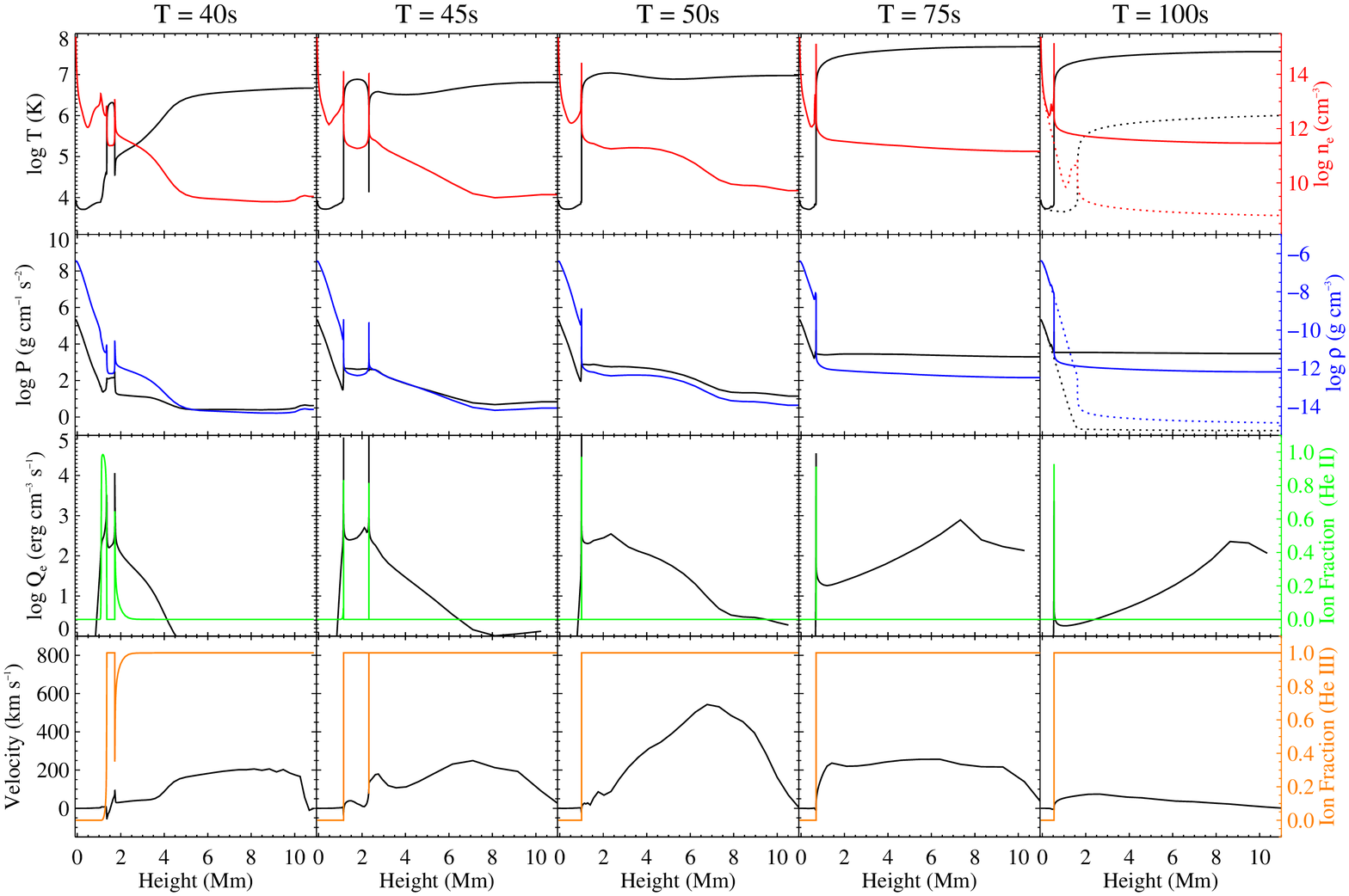}
\caption{Structure and evolution of the simulated solar atmosphere. The top panels display the atmosphere in the early phase when the beam-heating rate is low. The bottom panels show the evolution of the atmosphere during the explosive phase. The temperature (black) and electron density (red) are shown in the top row. The second row shows the pressure (black) and mass density (blue). The third row displays the beam-heating rate (black) and ionisation fractions of He II (green). The bulk velocity of the atmosphere (black) and ionisation fraction of He III (orange) are shown in the bottom row. The original atmosphere is overplotted as dotted lines at T=100s to illustrate the change in atmospheric structure.}
\label{f5}
\end{figure*}

\section{Model}\label{s:model}

The derived electron beam parameters were used as input to the one-dimensional NLTE radiative transfer and hydrodynamics modelling code RADYN. In the code, the plane-parallel equations of hydrodynamics are solved along with the statistical equilibrium, population conservation, and radiative transfer equations on an adaptive grid \citep{dorf87} that consists of 191 grid points with a base in the photosphere and an apex in the corona at a height of 10 Mm. Solutions are obtained for a six-level hydrogen atom, a nine-level helium atom, a six-level singly ionised calcium atom, and a four-level singly ionised magnesium atom. In total, 24 bound-bound and 22 bound-free transitions are solved in detail, including the first four transitions of the Lyman series, the first three Balmer transitions, the \ion{Ca}{II} H \& K lines and infrared triplet, the \ion{Mg}{II} h \&\ k lines, and the \ion{He}{II} 304 \AA\ line. Complete frequency re-distribution was assumed for all transitions except the Lyman transitions, for which the effects of partial re-distribution are mimicked by truncating the line profiles at ten Doppler widths \citep{milk73}. The equations are solved using five angle points and up to 100 frequency points for each transition. Other atomic species are included in the calculations as background continua in LTE using the Uppsala opacity package of \citet{gust73}. Optically thin cooling from bremsstrahlung and coronal metals is included using emissivities from the CHIANTI atomic database \citep{dere97,land13}. The beam-heating function of \citet{abbe99} based on the analytical treatment of \citet{emsl78} was used for an electron distribution described by a single power law. Non-thermal collisional excitation and ionisation was included for hydrogen using the treatment of \citet{fang93}. A complete description of the method of solution can be found in \citet{abbe99} and \citet{allr05}. The simulation was performed on 12 processor cores, and the model presented here took approximately one month to complete the calculations. Electron beam heating ended after 110 seconds in the simulation, then the model was allowed to continue evolving for a further 300 seconds.
        
\subsection{Atmosphere structure and evolution}\label{ss:ats}

The evolution of the model is plotted in Fig. \ref{f5}, displaying the loop temperature, electron density, gas pressure, mass density, beam heating rate, velocity, and ionisation fractions of \ion{He}{II} and \ion{He}{III} as a function of height in the atmosphere at ten times during the beam-heating phase of the simulation. The initial pre-flare atmosphere is structured with a chromosphere from 1 - 1.6 Mm above the photosphere (defined as the height where $\tau_{500nm} = 1$), the transition region (TR) at 1.6 Mm, and a corona with a temperature of 1 MK from 2 - 10 Mm (Fig. \ref{f5}, top left panel). The location of highest non-thermal electron energy deposition was approximately 1.0 Mm above the photosphere at $t=0 s$, at the base of the pre-flare chromosphere. The energy deposition rate for the first 30s was approximately 10$^{10}$ ergs cm$^{-2}$ s$^{-1}$ and the initial phase of the flare was computed quickly as the atmosphere remained in the regime of gentle evaporation, with the majority of the energy input by the beam being balanced by optically thin radiative losses. This initial heating resulted in an upward motion of the transition region and an extended region of increased temperature and electron density between 1 - 3 Mm. 

The energy deposition rate into the atmosphere began to increase after 23:20:00 UT ($t = 30$ s) and reached $10^{11}$ erg cm$^{-2}$ s$^{-1}$ at approximately 23:20:10 UT, at which time the explosive evaporation phase began. In the region of peak beam heating, helium had become fully ionised by $t=$ 40s, the temperature of the plasma began to rapidly increase and a region of high pressure formed. Material expanded outwards from this region as a result of the pressure gradient between the cool material below and the low-density coronal material above \citep{fish85a}. A downward-directed compression wave moved through the atmosphere with a locally cool region referred to as the ``chromospheric condensation'' \citep{fish85c} located between the compression wave and the flare TR. The downward-directed wave propagated with velocities of a few 10's km s$^{-1}$, with decreasing velocity as it travelled into regions of higher density, while the explosively evaporated material propagated with velocities of several hundred km s$^{-1}$. By 45 s into the simulation, or approximately 5 s after the explosive phase began, the temperature of the lower atmosphere between 1.5 - 2 Mm had increased to 10 MK and the electron density in this region was $10^{11}$ cm$^{-3}$. This is illustrated in the bottom panels of Fig. \ref{f5} from T = 40 to 50 s. The presence of high-temperature dense plasma located at low altitudes in the simulation agrees with several observational studies of the temperature and density of footpoint plasma made during the impulsive phase, for instance \citet{huds94} and \citet{grah13}.

\begin{figure*}
\centering
\includegraphics[width=0.9\linewidth]{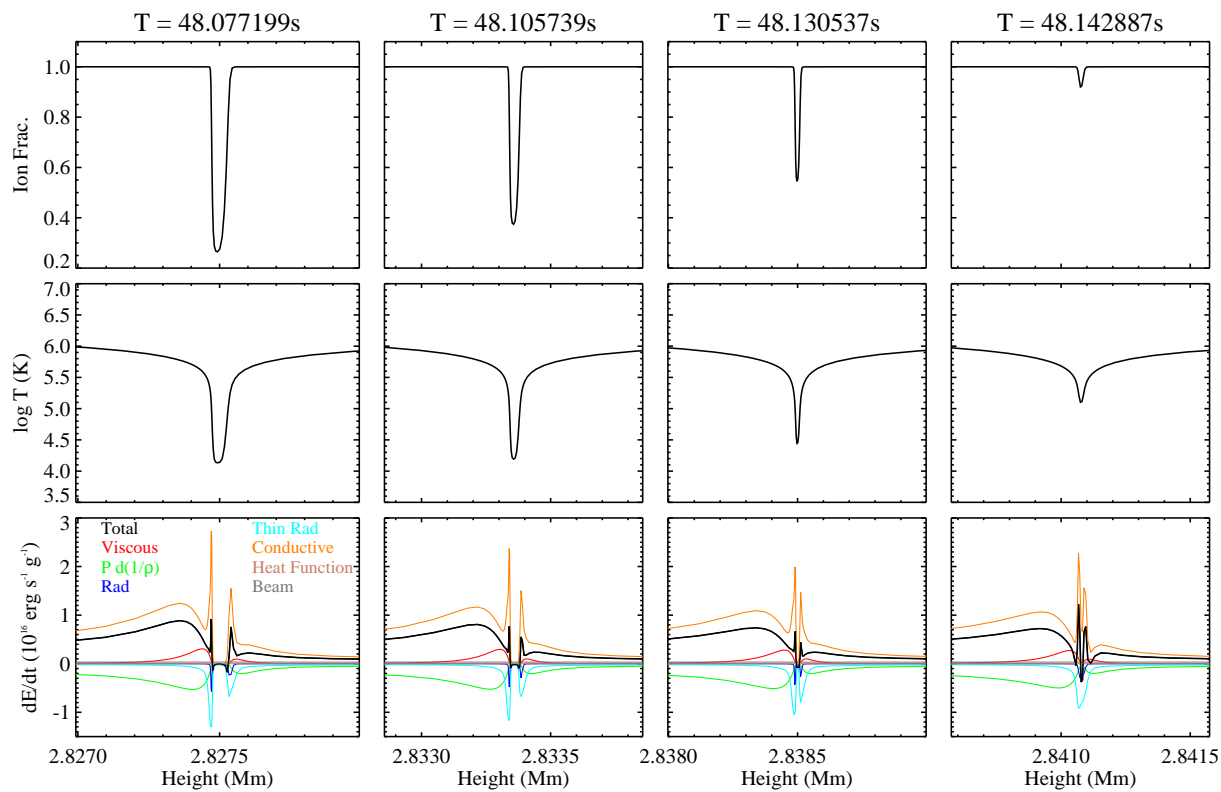}
\caption{Top panels: The ionisation fraction of He III from 48.07 to 48.14 s. Time is increasing from the left to right side of the plots. The middle panels display the temperature of the atmosphere. The bottom panels display the energy balance of the atmosphere. The quantities displayed are the net energy gain (black), viscous heating (red), work done by pressure (green), radiative losses from transitions solved in radiative transfer (blue), optically thin losses (cyan), conduction (orange), the background-heating function (brown), and the energy input from beam electrons (grey).}
\label{f6}
\end{figure*}

The expansion of the hot material outwards resulted in a very narrow, dense region moving upward through the atmosphere. The very high electron densities ($n_{e} = 10^{14}$ cm$^{-3}$) lead to a local temperature minimum due to the strong radiative losses from this material. However, after approximately 48 s of the simulation this region began to rapidly increase in temperature, from 10$^{5}$ K to over 10$^{7}$ K in approximately 0.1s. This is illustrated in Fig. \ref{f6}, which displays the ion fraction of \ion{He}{III}, the temperature, and energy balance of the atmosphere. Examining the energy balance of the atmosphere showed that there was a strong conductive energy flux (Fig. \ref{f6}, orange lines) into the cool material as a result of the steep temperature gradient. Radiative losses from optically thin emission (cyan lines) and from transitions solved in radiative transfer (blue lines) were insufficient to offset the energy input into this region, resulting in a net energy gain (black lines). This scenario appears similar to the initial explosive evaporation scenario that occurs in the chromosphere, where a high-energy flux leads to runaway ionisation and heating of the plasma. To our knowledge, this scenario has not been previously observed in RHD flare simulations, but the heating of this locally cool region has been observed in hydrodynamic flare simulations \citep[e.g.][]{naga84}. The very soft spectral indices and low values of E$_{c}$ that have been employed may be responsible for this because a greater proportion of non-thermal energy would be deposited higher in the corona compared to previous RHD simulations, which typically used harder electron distributions. As the temperature of the atmosphere in this cool region increased, the problem of solving the radiative transfer in the atmosphere was simplified as the transitions solved in radiative transfer were no longer being formed in two distinct regions of the atmosphere. Approximately 1.6 $\times 10^{6}$ time steps were required to reach 48.1s of simulated solar time, while only a further 1.0 $\times 10^{5}$ time steps were required to advance the simulation to 400s. The evolution of the atmosphere could then be studied over longer timescales compared to previous RHD flare simulations that were subjected to high-energy fluxes. We note that a similar atmosphere evolution has been observed in subsequent simulations involving very soft spectral indices and/or low values of $E_{c}$. A thorough investigation into the conditions that lead to this, involving a grid of models with varying spectral index, low-energy cutoff, and energy deposition rate, is currently being undertaken.

\begin{figure}
\centering
\includegraphics[width=0.9\linewidth]{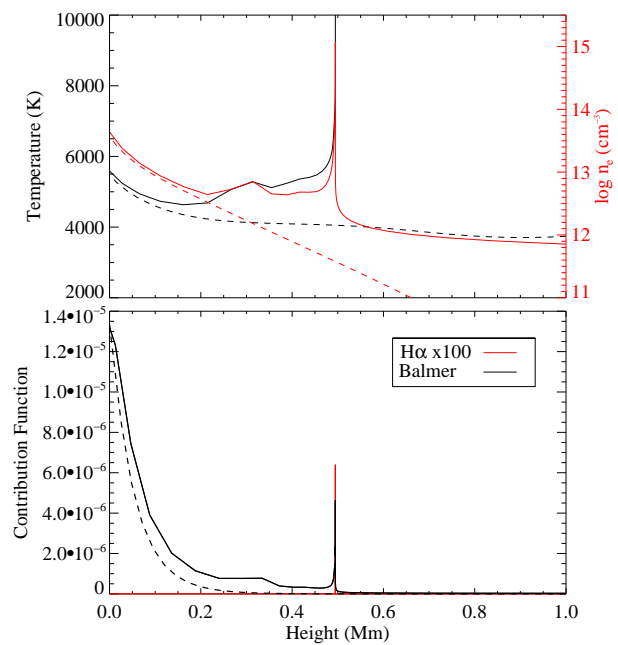}
\caption{Top panel: Structure of the lower atmopshere at 110s into the flare simulation. Solid black and red lines display the temperature and electron density. The dashed lines display the original atmosphere at t = 0s. Bottom panel: contribution function for the Balmer continuum at 3646 \AA\ and for H $\alpha$ at line centre (6564.7 \AA) for a viewing angle of $\mu =1$. The H $\alpha$ function was scaled by a factor of 100 to be visible on the same scale. The dashed black line displays the inital contributon function for the continuum at t = 0 s.}
\label{f7}
\end{figure}

As the simulation progressed, the peak of non-thermal electron energy deposition moved to higher altitudes as the coronal density increased and by 75s, the majority of energy was being deposited above 6 Mm. The deposition of energy in the corona produced a large increase in temperature. By the end of beam heating, the coronal temperature had increased from 1 to 50 MK and the mass and electron density had increased by over three orders of magnitude. The TR moved downward in altitude and the lower atmosphere was compressed. This is illustrated in the upper panel of Fig. \ref{f7}, which displays the temperature and density structure of the lower atmosphere between 0 - 1 Mm at 110 s. At this time, the flare TR was located at an altitude of 500 km above the photosphere. The temperature of the upper photosphere between 0.1 - 0.4 Mm increased by several 100 to 1000 K over the course of the simulation. By examining the contribution functions of the transitions solved in radiative transfer, we could investigate the formation heights of the different emission mechanisms. The intensity contribution function \citep{carl97} is given by Eq. \ref{eq:ci} and represents the fractional contribution to the emergent intensity originating from each height in the atmosphere. The terms in this equation are the source function $S_{\nu}$, the monochromatic opacity per unit volume $\chi_{\nu}$ , and the optical depth $\tau_{\nu}$ specified at frequency $\nu$ and a viewing angle $\mu$,
\begin{equation}
C_{i} = S_{\nu} \frac{\chi_{\nu}}{\tau_{\nu}} \tau_{\nu} e^{(-\tau_{\nu} / \mu)}
\label{eq:ci}
.\end{equation}
The bound-bound transitions were formed in an extremely narrow (<1 km) and high-density ($n_{e} \approx 10^{15}$ cm$^{-3}$) region at the base of flare TR, illustrated by the sharp peak in the H $\alpha$ contribution function in the bottom panel of Fig. \ref{f7}. The emitting location of these transitions changed as a function of time through the simulation as the TR moved downward through the atmosphere. The contribution function for the Balmer continuum at 3646 \AA\ is also plotted in this figure, illustrating that the optical continuum emission mainly originated from the heated layers of the upper photosphere with a local maximum across the flare TR.

Once energy deposition by non-thermal electrons ceased, the atmosphere began to cool and material drained from the corona. The energy balance of the atmosphere showed that conductive cooling dominated the coronal energy losses for the first 50 s following the end of beam heating, after which radiative cooling dominated the coronal energy loss. After 300s of simulated time, or approximately 200s after the end of beam heating, the coronal plasma had cooled down to 1 MK and then became radiatively unstable. The material present in the corona rapidly decreased in temperature, resulting in a cool (T $\approx 10^{4}$ K), dense ($n_{e} \approx 10^{11}$ cm$^{-3}$) loop.

\subsection{Electron collisional stopping depth}\label{s:estop}

\begin{figure}
\centering
\includegraphics[width=1.0\linewidth]{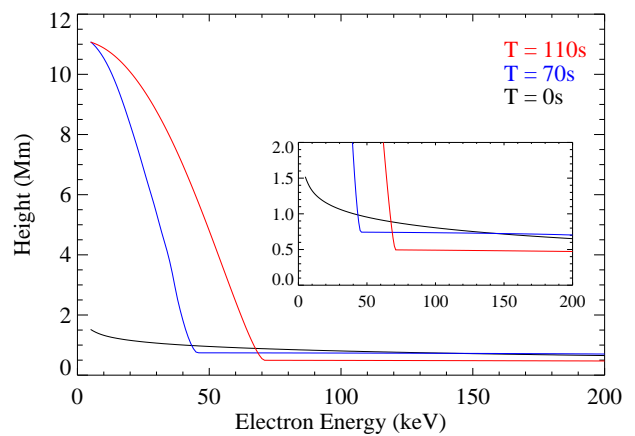}
\caption{Greatest penetration depth of electrons injected into the atmosphere as a function of energy from 5 - 200 keV for three times in the simulation. The inset figure displays a zoom-in of the atmosphere between 0 - 2 Mm.}
\label{f8}
\end{figure}

The changes in the mass structure of the atmosphere during the flare affect where non-thermal electrons will be collisionally stopped and lose their energy. We investigated where in the atmosphere electrons would be stopped at different times in the evolution of the simulated flare. The collisional stopping column density of electrons with energies from 5 to 200 keV was calculated using the analytical treatment of \citet{emsl78} (Eq. \ref{eq:ecden}), and the corresponding height in the atmosphere was then determined from the flare simulation. The results are shown in Fig. \ref{f8} for three times in the simulation at $t$ = 0, 70 and 110 s,
\begin{equation}
N_{e} =  \frac{E_{0}^{2}}{4 \pi e^{4} \Lambda} \approx 10^{17} E^{2} \:\:\:\: \text{[cm$^{-2}$}]
\label{eq:ecden}
.\end{equation}
The black line in Fig. \ref{f8} shows that the stopping height in the initial atmosphere for the energy range considered lies between 0.8  to 1.5 Mm above the photosphere. As the density of the corona increases, lower energy electrons are stopped higher in the atmosphere, with the majority of electrons being stopped near the apex of the coronal loop. If we examine the stopping depth of electrons of a specific energy, for example 50 keV, we can show how the greatest penetration depth changes throughout the simulation. In the initial atmosphere, 50 keV electrons would be stopped at approximately 1.1 Mm above the photosphere, while at 70s they can penetrate to around 0.75 Mm, as the column density required to stop the 50 keV electrons is now located at a lower altitude in the atmosphere. At the end of beam heating, the 50 keV electrons would now be able to penetrate no farther than 5 Mm into the atmosphere, as the density of the corona has increased. This illustrates that the column density required to collisionally stop an electron corresponds to different altitudes above the photosphere at different times in the simulation. The greatest possible penetration depth was approximately 0.45 Mm above the photosphere at 130s into the simulation for electrons of energy 70 keV and higher. They were stopped at the base of the flare TR where there is a sharp increase in the column density to over $10^{22}$ cm$^{-2}$. These low altitudes are similar to WL observations made in the continuum at 6173 \AA\ of a limb flare by \cite{mart12}, who found WL source heights of $\approx$ 0.3 Mm above the photosphere and co-spatial with 30 - 80 keV HXR emission. These source heights were located below the expected penetration depth of the 50 keV electrons that would be required to produce the observed HXR emission. The downward mass motion and compression of the lower atmosphere in our simulation could offer an explanation for the co-spatial WL and HXR emission present at low altitudes in this observed flare.

From the analytical formula for the electron collisional stopping column density we have demonstrated that it is possible for energetic electrons to propagate farther into a flaring atmosphere compared to a quiet-Sun atmosphere. This is explained by material being compressed below the altitude at which explosive evaporation occurs. In general, it was found that electrons with energies below 50 keV are collisionally stopped at a higher altitude as material is ablated into the corona, while the more energetic electrons ($>$60 keV) are able to propagate farther into the atmosphere. It required continuous energy input into the atmosphere over a timescale of 100 s to produce these deep penetration depths. The simulation presented here used very soft spectral indices and moderate energy deposition rates. The use of higher energy deposition rates may produce a more rapid compression of the atmosphere, allowing electrons to penetrate to low altitudes on shorter timescales. Finally, we note that this analysis is quite simplified, using the analytical expressions for electrons penetrating into a fully ionised cold-target. A more advanced treatment using the Fokker-Planck version of the RADYN code may alter the depth at which the electrons are able to penetrate to, as it includes pitch-angle scattering and magnetic mirroring.

\section{Observational comparison}\label{s:comp}

The results of the model, including hydrodynamic variables and radiative emissions, were compared to observations made with EVE, GOES, and RHESSI. A pre-flare background averaged over a five-minute period was subtracted from the EVE data to obtain the flare irradiance.

\subsection{Hydrodynamic variables}\label{ss:hyd}

The evolution of the hydrodynamic variables during the studied flare were measured using spectroscopic observations from the EVE MEGS-A instrument. The \ion{Fe}{XXI} 145/128 density-sensitive line ratio was used to determine the mean electron density of the plasma \citep{mill12b}. Emission lines from \ion{Fe}{VIII} to \ion{Fe}{XXIV} allowed for the temperature structure of the flaring plasma to be estimated from a differential emission measure (DEM) analysis \citep{kenn13}. The DEMs were constructed using coronal abundances \citep{feld92} over the temperature range log T$_{e}$ = 6.0 - 7.5. Estimates of the isothermal T and EM were obtained from GOES filter ratios \citep{whit05} available through the \textit{goes} widget in SolarSoft, and the T and EM were also determined from analysis of RHESSI spectra. The evolution of the electron density, T and EM were measured and compared to the modelled parameters. Detailed descriptions of the diagnostics methods can be found in the papers referenced above. The model coronal electron density was calculated in a manner to directly compare with the mean electron density determined from  \ion{Fe}{XXI} density-sensitive line ratios. An average value was calculated using the grid cells that fell within the typical formation temperature range of \ion{Fe}{XXI} (log T$_{e}$ = 6.8 - 7.2). The value of $n_{e}$ at each grid cell was weighted by the height $\Delta z$ of that cell, to avoid the average density being dominated by regions of the highest spatial resolution. 

\begin{figure}
\centering
\includegraphics[width=1.0\linewidth]{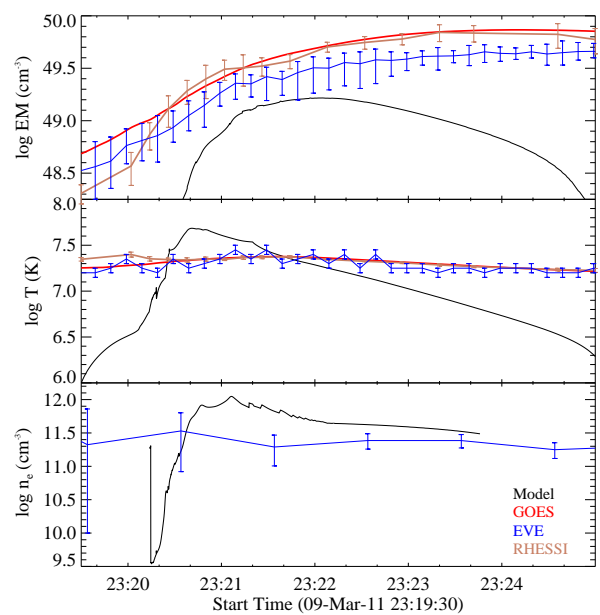} 
\caption{Comparison of the modelled and observed emission measure, temperature, and electron density. The red lines are the measurements from GOES, blue lines from EVE, brown lines from RHESSI, and black lines are the modelled values.}
\label{f9}
\end{figure}

Figure \ref{f9} displays the modelled and observed values of the EM, T, and electron density. The temperature measurement from EVE is the peak temperature of the DEM at each time bin, and the EM is the integral of the DEM over the range log T$_{e}$ = 6.0 - 7.5. Values of the flare emission measure obtained from RHESSI and GOES are very similar. However, the EM derived from EVE is consistently lower than both of the RHESSI and GOES measurements. The model EM was the integral of the model DEM above log T$_{e} \ge 6.0$. It was scaled by an estimate of the emitting area and has a peak value of log EM = 49.1 after 150 s into the simulation. The peak temperature determined from all three instruments was approximately log T$_{e}$ = 7.3 (20 MK), while the peak loop temperature of the model reached log T$_{e}$ = 7.7 (50 MK). The combined EVE and RHESSI DEM for this event was studied by \citet{casp14}, who also found peak temperatures between 20 - 30 MK with no evidence for a super-hot component. These observations indicate that the coronal temperature in the model is too high. A possible explanation for this is that the value of $E_{c}$ determined from the spectral fitting was too low, resulting in a large amount of energy being contained in lower energy electrons that deposit their energy higher in the atmosphere. Alternatively, it might mean that the continuous deposition of energy into a single model atmosphere for a prolonged period of time is unrealistic. The comparison with the observed electron density was limited as the density-sensitive lines were weak in the EVE spectra resulting in large irradiance uncertainties, and were also affected by line blending. The modelled \ion{Fe}{XXI} electron density reached a peak value of log n$_{e}$ = 11.8 after 100s of the simulation followed by a slow decay over several minutes. EVE measurements indicated values of log n$_{e}$ = 11.0 - 11.5, which were slightly lower than the model atmosphere. However, the density at the apex of the modelled loop (see Fig. \ref{f5}) agreed better with the observed electron density, reaching a peak value of approximately  log n$_{e}$ = 11.5. 

The temporal evolution of the modelled parameters was also quite different from what was observed. Pre-impulsive phase heating leads to high-temperature plasma being present in the corona before a detectable increase in HXR emission, and the temperature and EM decrease much more rapidly than in the observations. Additional energy release in the decay phase that has not been accounted for in our simulation could explain this. This would agree with previous numerical modelling and observational studies of decay phase cooling, which indicated that additional energy release is required to explain the cooling times of flare loops, for instance \citet{reev02} and \citet{ryan13}. It is possible to add a time-dependent volumetric heating rate to the simulation to account for additional energy release in the decay phase of the flare. This would lead to a more gradual decay as the additional heating would offset the radiative losses from the atmosphere. However, we have not attempted to model this without a physical description of the additional energy deposition in the atmosphere.

\begin{figure}[h]
\centering
\includegraphics[width=0.9\linewidth]{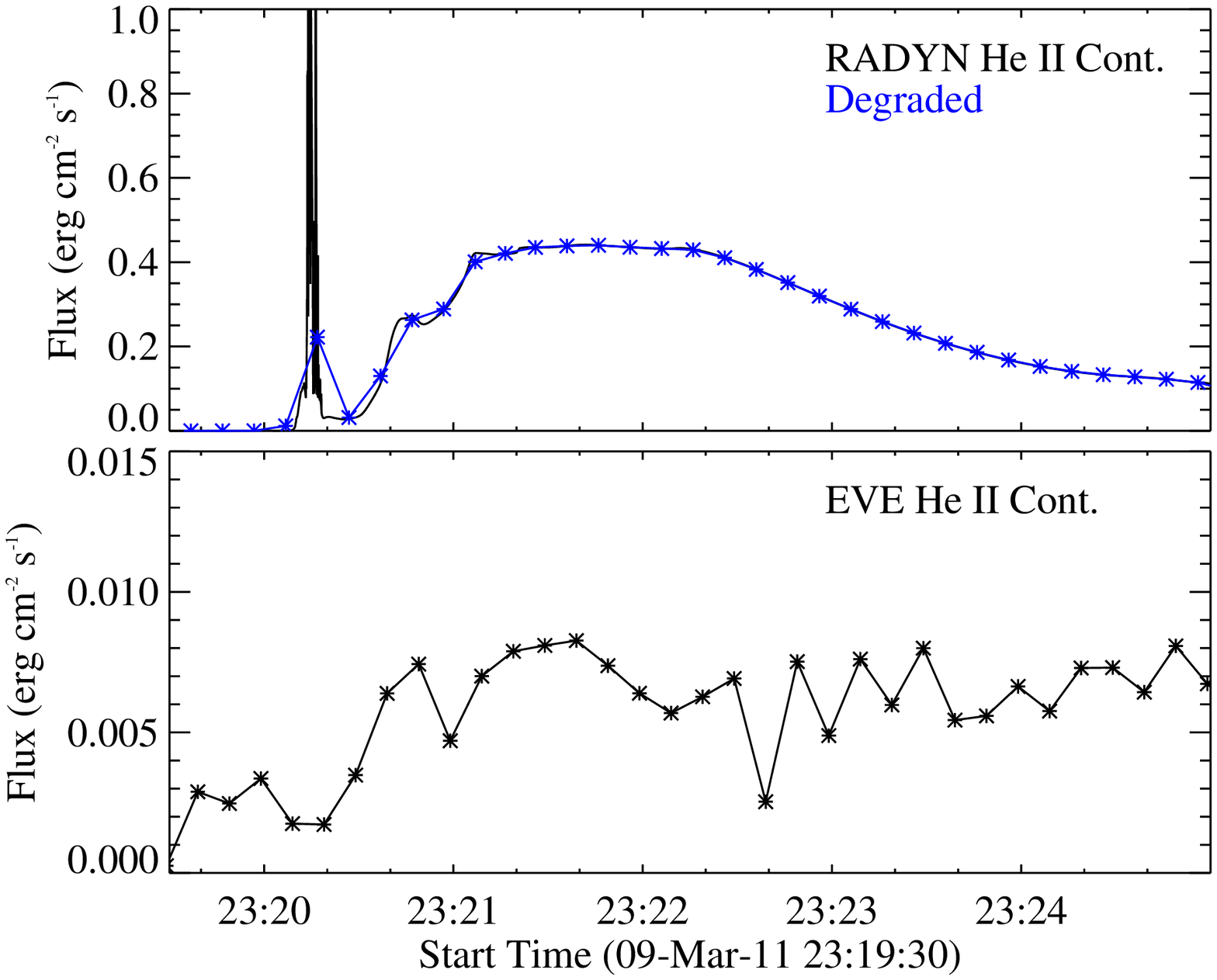} 
\includegraphics[width=0.9\linewidth]{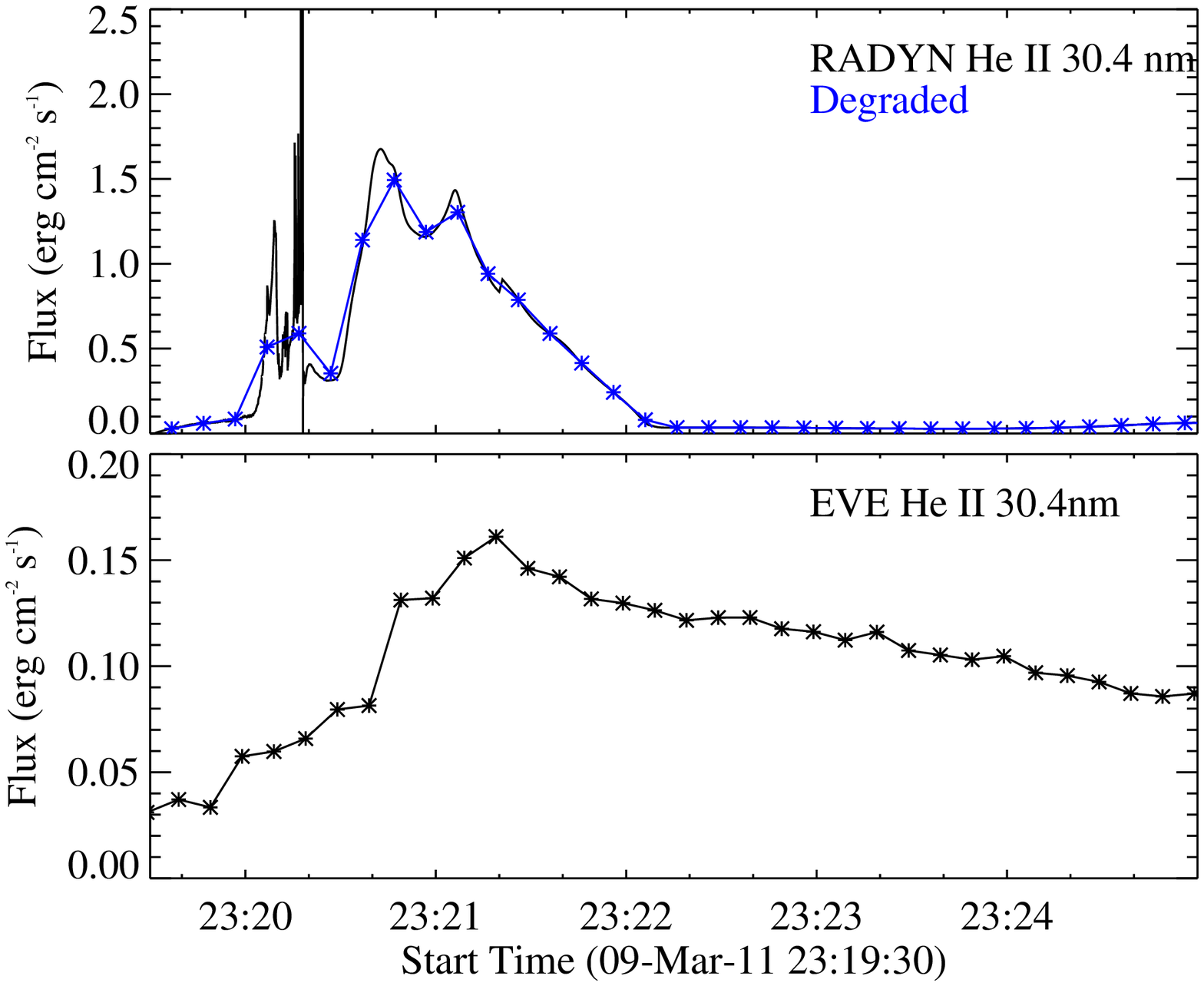} 
\caption{Light curves of \ion{He}{II} recombination and 304 \AA\ line emission created from the RADYN model compared to observed light curves from EVE. The modelled light curves are displayed at each saved time step and then degraded to the instrumental cadence and exposure time of EVE (plotted in blue).  The RADYN light curves have been scaled by an estimate of the flare area and converted to the flux measured at Earth to facilitate direct comparison with the observations. The observed light
curves from EVE MEGS-A are displayed at 10s cadence.}
\label{f10}
\end{figure}

\subsection{Light curves and synthetic spectra}\label{ss:lcurves}

The only bound-bound transition that is solved in NLTE radiative transfer in RADYN and was observed during this event is the \ion{He}{II} 304 \AA\ doublet. This transition and the \ion{He}{II} recombination continuum are observed by EVE MEGS-A. The flux of the 304 \AA\ doublet was measured by fitting with a Gaussian profile, while the flux of the continuum was measured using a method similar to that described by \citet{mill12a}, by fitting with an exponential function and subtracting the underlying free-free emission. Light curves of EUV emission lines and the \ion{He}{II} doublet were compared to synthetic light curves generated from the RADYN model. The RADYN light curves are displayed using all output time steps, but to facilitate direct comparisons with observations, the light curves are also displayed degraded to the cadence of the observing instrument. Flux was averaged over the instrument exposure time and sampled at the instrument cadence (both 10 seconds for EVE). 

The modelled and observed light curves of the \ion{He}{II} recombination continuum and \ion{He}{II} 304 \AA\ are shown in Fig. \ref{f10}. The degraded light curve is plotted in blue. The modelled values are for a viewing angle of $\mu = 0.95$ and have been scaled by an estimate of the flare area then converted into flux that would be measured at Earth to facilitate comparison with the observations. The observed light curves are displayed below each of the modelled light curves. In the first 45s of the simulation the light curves feature sharp, sub-second variations in intensity. These variations are smoothed when the light curve is degraded to the instrumental exposure time. After the cool material moving upward through the corona increased in temperature, the intensity of \ion{He}{II} declined to a minimum as helium became fully ionised in this region. The intensity then increased to a maximum as the density in the region of line formation at the base of the TR peaked at approximately 75s. The time of peak intensity of the observed and modelled \ion{He}{II} 304 \AA\ light curves occurred to within 40s of each other. The modelled light curve decayed faster as energy input into the atmosphere ceased after 100s of the simulation. The modelled 304 \AA\ emission was an order of magnitude greater than the measured flux of the line by EVE, and the modelled continuum was approximately 2 orders of magnitude greater than the observed continuum flux. 

\begin{figure}
\centering
\includegraphics[width=0.9\linewidth]{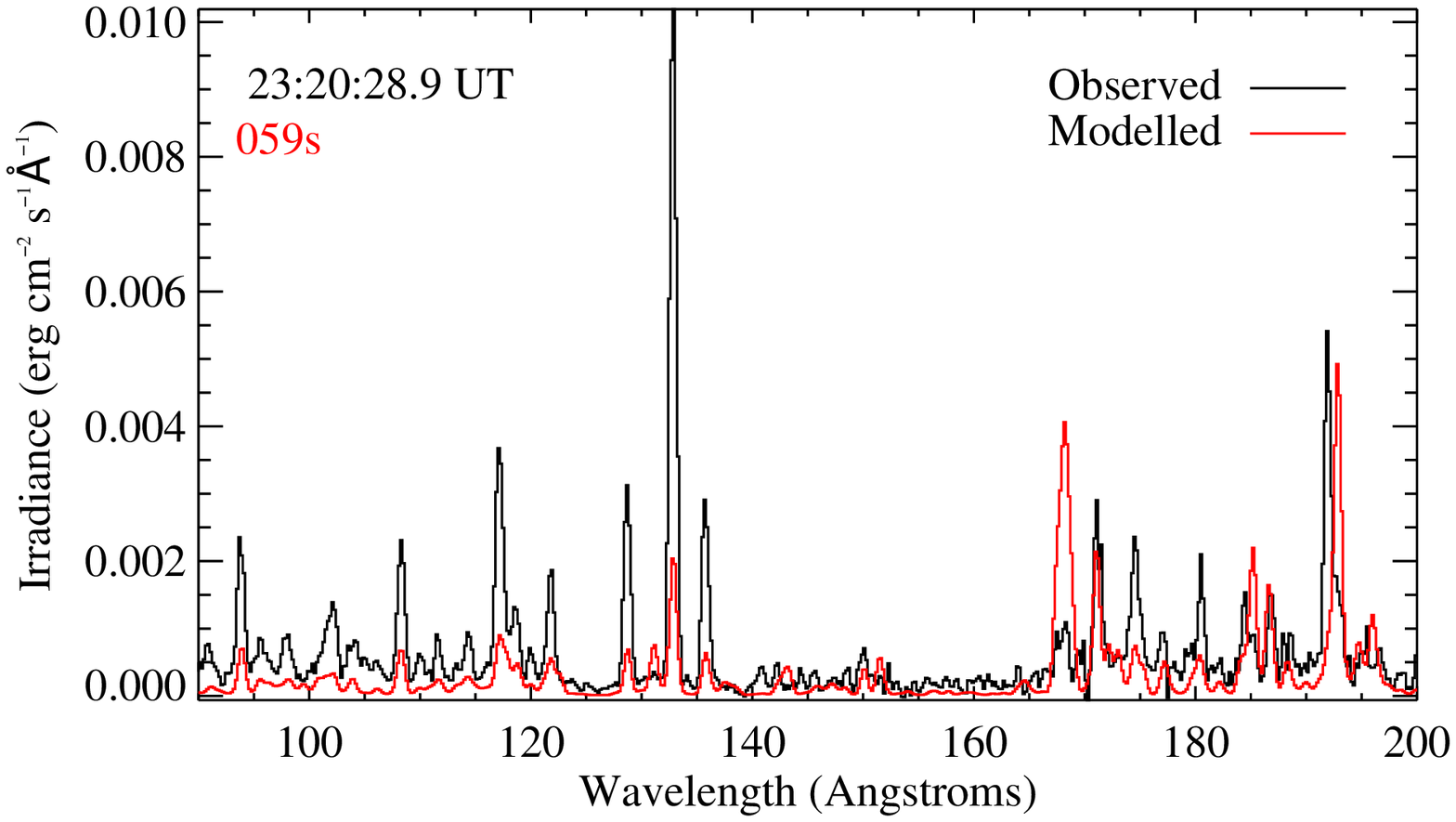} 
\includegraphics[width=0.9\linewidth]{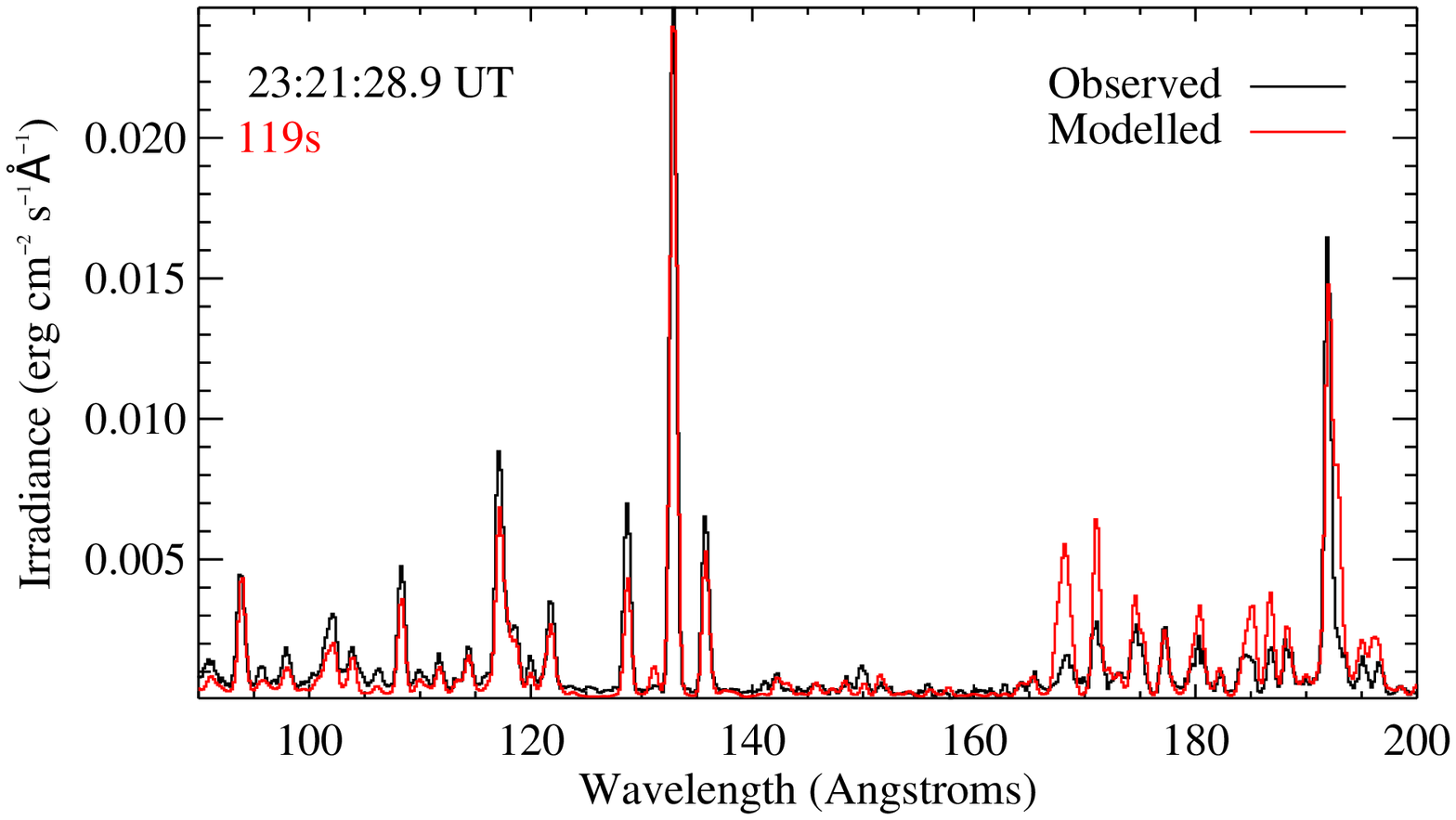} 
\includegraphics[width=0.9\linewidth]{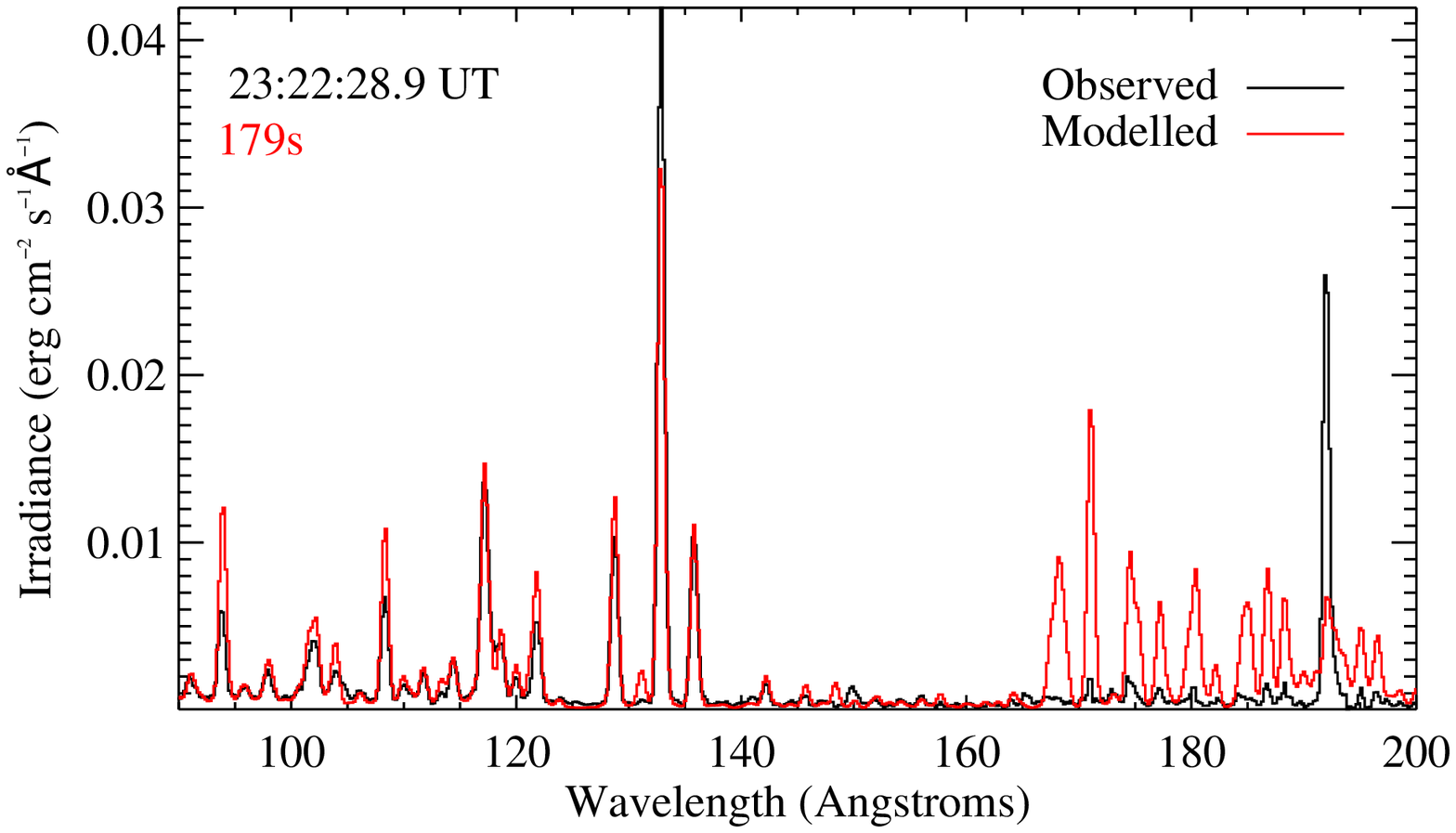} 
\caption{Observed (black) and synthetic (red) spectra of the 90 to 200 \AA\ wavelength range. The prominent emission lines are the flare lines of \ion{Fe}{XVIII} to \ion{Fe}{XXIII} from 90 - 150 \AA, emission lines of \ion{Fe}{VIII} to \ion{Fe}{X} near 165 - 180 \AA,\ and the \ion{Fe}{XXIV} line at 192 \AA.}
\label{f11}
\end{figure}

\begin{figure}
\centering
\includegraphics[width=0.95\linewidth]{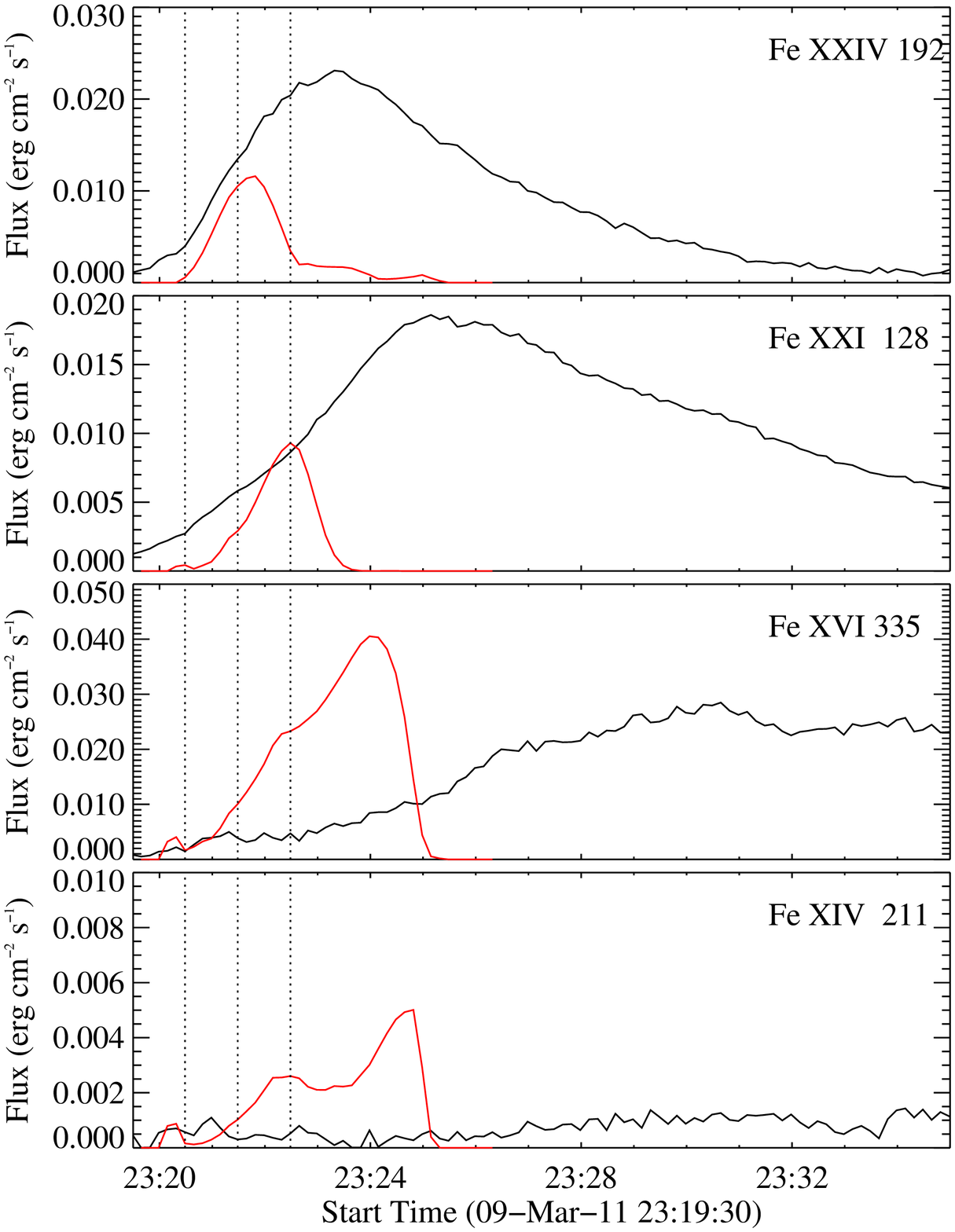} 
\caption{Observed (black) and synthetic (red) light curves of emission lines from Fe ions observed by EVE. The dotted lines indicate the times corresponding to the spectra displayed in Fig. \ref{f11}.}
\label{f12}
\end{figure}

Synthetic EUV spectra were generated by constructing DEMs of the simulated atmosphere averaged over 10 s intervals. The CHIANTI atomic database \citep{land13} was then employed to generate spectra of the 60 - 370 \AA\ wavelength range at the wavelength binning (0.2 \AA) of the EVE MEGS-A instrument and at the full width half maximum of observed lines in EVE spectra (0.7 \AA), assuming coronal abundances \citep{feld92}. The comparison between the observed and synthetic EUV emission is displayed in Fig. \ref{f11} at three times during the flare evolution. The best agreement between the modelled and observed EUV emission is found at approximately 100 - 140 s into the simulation (Fig. \ref{f11}, middle panel). At these times, the modelled irradiance agreed with the observed irradiance to within a factor of two. Considering that the calculation of the synthetic irradiance relied on the assumption of ionisation equilibrium, elemental abundances, and an estimate of flare area, this is reasonably good agreement. The irradiance of emission lines formed at temperatures of between 0.1 and 1 MK was overestimated by around a factor of ten during the impulsive phase and peak of the flare. The overestimation of the intensity of UV emission lines was also found in the models of \citet{fish85b}. This is due to the high electron densities across the model flare TR, which resulted in an extremely high EM at these temperatures, despite the narrow grid cell heights. During the gradual phase coronal dimming \citep{rust76, wood11} affected the spatially integrated EVE observations and reduced the observed irradiance of these lines, preventing an accurate comparison to the modelled irradiance. Light curves of several emission lines were generated by fitting the synthetic spectra with Gaussian profiles in the same manner as observed EVE spectra are analysed. The line irradiance and emission profiles were compared to what was observed during the flare (see Fig. \ref{f12}). Emission from \ion{Fe}{XXIV} line peaked first at 130s into the simulation with emission from cooler lines peaking later as the corona cooled. The \ion{Fe}{XXIV} emission peaked approximately 100s before the observed light curve, with a longer delay between the time of observed and modelled peak flux for cooler emission lines. At 330s into the simulation, the loop became radiatively unstable and cooled to approximately 10$^{4}$ K.

\section{Conclusions}\label{s:conclu}

We have simulated the response of the solar atmosphere to non-thermal electron beam heating using a 1D RHD modelling code. Our aim was to compare the modelled and observed flare parameters and line irradiance and to investigate the structure of the lower atmosphere during the flare. During the explosive evaporation phase the region of cool, dense material moving upward through the atmosphere rapidly increased in temperature. Despite the very high electron densities, radiative losses were unable to balance the energy input into this region via thermal conduction. The very soft spectral indices and low values of $E_{c}$ that were employed could explain this, due to the greater proportion of non-thermal electron energy that would be deposited higher in the atmosphere. This greatly simplified the problem of solving the radiative transfer in the atmosphere and allowed the evolution of the atmosphere to be studied over longer timescales compared to previous RHD flare simulations involving very high energy deposition rates.

It was found that the structure of the lower atmosphere became extremely compressed during the simulation. The flare TR moved downwards in altitude during the explosive phase and was located at 440 km above the photosphere at 120 s. By calculating the collisional stopping depth of electrons, it was determined that the downward mass motion and compression of the lower atmosphere allowed for higher energy electrons to penetrate to a lower altitude above the photosphere when compared to the pre-flare atmosphere. The location of deepest penetration coincides with the region where line emission originates and with a peak in the intensity contribution function of optical continua. This could help to explain observations of low-altitude, co-spatial WL and HXR emission \citep{mart12}, which was considered contradictory to the standard thick-target model. However, continuous heating of the atmosphere over a timescale of 100s was required to produce these electron penetration depths.

High-temperature plasma was produced at the footpoint of the modelled loop during the impulsive phase. Less than 10s after explosive evaporation occurred the temperature in the lower atmosphere increased to 10 MK at altitudes of 1 - 2 Mm. The electron density in this region was approximately 10$^{11}$ cm$^{-3}$. This generally agrees with observational studies of the temperature and density of flare ribbons and footpoints, which indicate the presence of high-temperature material present at low altitudes in the solar atmosphere. The model produced a denser and hotter flare loop than what was detected and the temporal evolution of the loop temperature, electron density, and EM did not agree with observations. The modelled loop cooled more rapidly than what was observed, suggesting additional energy release in the decay phase. The comparison between observed and synthetic EUV irradiance indicated good agreement during the peak of the simulated flare, to within a factor of 2 for the majority of emission lines. However, the irradiance of lower temperature emission lines from ions formed at temperatures lower than 10$^{6}$ K was significantly overestimated during the very early and the decay phase of the simulated flare.

Although the energy deposition rate into the atmosphere is an underestimate, modelling of this flare has highlighted several interesting results. A grid of flare models is currently being generated to investigate the atmospheric response to heating from different distributions of non-thermal electron energy across the observed range of the parameters that describe the electron distribution. The inclusion of higher energy fluxes and the adoption of different pre-flare atmospheres should also be investigated to determine how this affects the chromospheric response to non-thermal energy input and the evolution of the flare atmosphere.

\begin{acknowledgements}
M.B.K. thanks the Northern Ireland Department of Employment and Learning for the award of a PhD studentship and acknowledges support from the International Space Science Institute to attend a team meeting led by L. Fletcher on "Observations and Modelling of Flare Chromospheres", where helpful discussions regarding the manuscript took place. R.O.M. is grateful to the Leverhulme Trust for financial support from grant F/00203/X, and to NASA for LWS/TR\&T grant NNX11AQ53G and LWS/SDO Data Analysis grant NNX14AE07G. M.M and F.P.K. acknowledge financial support from the UK STFC. The research leading to these results has received funding from the European Community's Seventh Framework Programme (FP7/2007-2013) under grant agreement no. 606862 (F-CHROMA). CHIANTI is a collaborative project involving George Mason University, the University of Michigan (USA) and the University of Cambridge (UK). Data are provided courtesy of NASA/SDO and RHESSI, and the EVE, HMI,  and RHESSI science teams. This research has made use of NASA's Astrophysics Data System. We thank the anonymous referee for their comments that helped to improve the quality and clarity of the manuscript. \end{acknowledgements}

\bibliographystyle{aa}

\end{document}